\documentclass[english]{article}
\usepackage{amsmath}
\usepackage{amssymb}
\usepackage{cite}
\usepackage{hyperref}
\usepackage{amsbsy}
\usepackage{amsfonts}
\usepackage{amsopn}
\usepackage{amstext}
\usepackage{graphicx}
\usepackage{amssymb}
\usepackage{amsfonts}
\usepackage{amsmath}
\usepackage{amsmath,amsthm,amsfonts,amssymb}
\usepackage[mathcal]{eucal}
\usepackage{mathrsfs}
\usepackage{graphicx}
\usepackage[english]{babel}
\usepackage{color}
\usepackage{esint}
\usepackage[dvips]{epsfig}
\usepackage[dvips]{graphicx}
\usepackage{float}
\usepackage{units}

\def\equ#1{(\ref{#1})}
\def\be#1{\begin{equation}\label{#1}}
           
       \def\ee{\end{equation}}

\makeatletter

\begin{document}
\begin{titlepage}
\thispagestyle{empty}

\bigskip

\begin{center}
\noindent{\Large \textbf
{The noncommutative quantum Hall effect with anomalous magnetic moment in three different relativistic scenarios}}\\

\vspace{0,5cm}

\noindent{R. R. S. Oliveira ${}^{a,}$\footnote{e-mail: rubensrso@fisica.ufc.br}, G. Alencar ${}^{a,}$\footnote{e-mail: geova@fisica.ufc.br} and R. R. Landim ${}^{a,}$\footnote{e-mail: renan@fisica.ufc.br}}

\vspace{0,5cm}

{\it ${}^a$Departamento de F\'{\i}sica, Universidade Federal do Cear\'{a} -
Caixa Postal 6030, Campus do Pici, 60455-760, Fortaleza, Cear\'{a}, Brazil. 
 }

\end{center}

\vspace{0.3cm}

\begin{abstract}

In the present paper, we investigate the bound-state solutions of the noncommutative quantum Hall effect (NCQHE) with anomalous magnetic moment (AMM) in three different relativistic scenarios, namely: the Minkowski spacetime (inertial flat case), the spinning cosmic string (CS) spacetime (inertial curved case), and the spinning CS spacetime with noninertial effects (noninertial curved case). In particular, in the first two scenarios, we have an inertial frame, while in the third, we have a rotating frame. With respect to bound-state solutions, we focus primarily on eigenfunctions (Dirac spinor and wave function) and on energy eigenvalues (Landau levels), where we use the flat and curved Dirac equation in polar coordinates to reach such solutions. However, unlike the literature, here we consider a CS with an angular momentum non-null and also the NC of the positions, and therefore, we seek a more general description for the QHE. Once the solutions are obtained, we discuss the influence of all parameters and physical quantities on relativistic energy levels. Finally, we analyze the nonrelativistic limit, and we also compared our problem with other works, where we verified that our results generalize some particular cases of the literature.

\end{abstract}

\end{titlepage}
\tableofcontents
\newpage
\section{Introduction}

The quantum Hall effect (QHE) \cite{Klitzing} is undoubtedly one of the most peculiar and impressive physical phenomena of quantum physics. One of the most striking aspects of the QHE is the fact that the energy spectrum (Landau levels), the electrical conductivity (Hall conductivity), and the electrical resistivity (Hall resistivity) of a two-dimensional electron gas (2DEG) are quantized quantities when subjected to low temperatures and strong magnetic fields. In experimental approaches, the QHE has been observed in graphene \cite{Kane,Novoselov}, quantum wells \cite{Bernevig}, and in semiconductors \cite{Kato}. Already in theoretical approaches, the QHE has been studied in nonrelativistic \cite{Yoshioka,Das,Dayi,Dulat} and relativistic \cite{Schakel,Haldane,Beneventano,Lamata} quantum mechanics (low and high energy regimes). In the literature, the QHE also is so-called incompressible quantum fluid \cite{Laughlin}, and can be manifest in many ways, such as in the integer QHE \cite{Klitzing,Gusynin}, fractional QHE \cite{Zhang2}, anomalous QHE \cite{Laughlin,Goldman}, and in the spin and orbital QHE \cite{Kane,Bernevig,Tanaka}, respectively. In addition, the QHE was already investigated in different contexts of physics, for example, in the effective-field-theory model \cite{Zhang2}, Hor\v ava-Lifshitz gravity \cite{Wu}, Chern-Simons-Landau-Ginzburg theory \cite{Z}, and in string theory \cite{Fabinger}. Currently, the QHE is used as the standard of electrical resistance by metrology laboratories around the world \cite{Jeckelmann}, and recently was studied in Rabi oscillations \cite{Tran}, Weyl semimetals \cite{Thakurathi}, non-Hermitian systems \cite{Yoshida}, parity anomaly \cite{Zou}, and in quantum field theory \cite{Kaplan}.

Since that noncommutative geometry (NCG) was introduced in quantum field theory \cite{Szabo,Douglas,Gomis} and in string theory \cite{Seiberg}, the study of quantum systems in NC spaces or NC phase spaces (general case) has been a subject of much interest and attention over the years \cite{Gamboa,Acatrinei,Bertolami,Berto}. The concept of NC spaces initially arose from a work of Snyder on quantized spacetimes \cite{Snyder} and is considered a possible scenario for the short-distance behavior (Planck length scale) of some physical theories \cite{Szabo,Majid,Abel,Pikovski}. Another motivation for the study of NC (phase) spaces is due to the supposed quantization of gravitational fields, where the NC may be, possibly, a result of the effects of quantum gravity \cite{Moffat}. For more details on the phenomenology of NCG, see Refs. \cite{Hinchliffe,Melic,Buric,Schupp,Abel,Pikovski}, where supposed signatures of NC were investigated in the decay of kaons and vector bosons, photon-neutrino interaction, vacuum birefringence, and in quantum optics. In essence, NC spaces (or NC spacetimes) are based on the assumption that the position operators are NC variables, i.e., satisfies $[x^{NC}_\mu,x^{NC}_\nu]\neq{0}$, while that the NC phase spaces are based on the assumption that the momentum operators also are NC variables, i.e., satisfies $[p^{NC}_\mu,p^{NC}_\nu]\neq{0}$, respectively. Besides, NC (phase) spaces are applied in several important areas of physics, such as in quantum chromodynamics \cite{Carlson}, quantum electrodynamics (QED) \cite{Riad}, black holes \cite{Nicolini}, standard model \cite{Melic,Buric}, quantum cosmology \cite{Garcia}, violation of Lorentz symmetry \cite{Berto}, Shannon entropy \cite{Nascimento}, graphene \cite{Bastos}, and in the QHE \cite{Dayi,Dulat}.

In nonrelativistic quantum mechanics, it is known that the orbital and spin magnetic dipole moment (MDM) originate from the nonrelativistic limit of the Dirac equation (DE) with minimal coupling \cite{Greiner}. However, when quantum electrodynamics (QED) effects are taken into account, a correction for to spin MDM is found, where such correction is given by the so-called anomalous magnetic moment (AMM), which is a dimensionless quantity that arises from Feynman diagrams with loops (for fermions interacting with strong external electromagnetic fields) \cite{Greiner1,Geiger,Greiner}. Indeed, one of the most important predictions of the QED was the AMM, which agrees with the experimental value by more than 10 significant digits \cite{Greiner1,Van}. On the other hand, unlike spin MDM, MDM that arises from AMM is not an intrinsic property of the fermion, but originates from the interaction (radiative self-interactions) with the external electromagnetic field \cite{Greiner1,Geiger,Greiner}. Furthermore, QED in NC spaces demonstrated that fermions have an intrinsic (anomalous) MDM regardless of the spin \cite{Riad}. In particular, relativistic and nonrelativistic Dirac fermions with AMM have already been investigated in different contexts of physics, such as in the supersymmetric standard model \cite{Moroi}, CPT and Lorentz symmetry \cite{Stadnik}, dark matter \cite{Barger}, light-by-light scattering \cite{Blum}, Aharonov-Casher effect \cite{Mirza,Hagen}, and NC quantum mechanics \cite{Passos,Ribeiro}, where recently were studied with electromagnetic and gravitational fields \cite{B1,B2,Dvornikov}, and its thermodynamic properties were also analyzed \cite{Rubens}.

In recent decades, inertial effects (here we call noninertial effects) generated by rotating frames (such as Coriolis, centrifugal, or Euler forces) on quantum systems have been widely studied in the literature, where possibly the oldest and most well-known effect on this theme is the Barnett effect (magnetization induced by rotation) \cite{Barnett,Hendricks,Ono}. In the nonrelativistic context (low energies), noninertial effects are very important in physical systems found in condensed matter (theoretical and experimental), for example, have already been applied to problems involving Bose-Einstein condensates \cite{Cooper}, spin currents \cite{Matsuo1,Matsuo2}, atomic gases \cite{Cooper2008}, fullerenes ($C_{60}$ molecules) \cite{Shen}, superconductors \cite{Fischer}, quantum rings \cite{Pereira}, and in the QHE \cite{Matsuo1,Filgueiras,Brand}. Now, in the relativistic context (high energies), noninertial effects are also very important \cite{Dv,Lambiase,Santos}, and considering in special the DE (or the Dirac field) in a rotating frame \cite{Hehl}, we can study systems involving boundary effects and gapped dispersion in rotating fermionic matter \cite{Ebihara}, chiral symmetry restoration, moment of inertia and thermodynamics \cite{Chernodub1}, chiral symmetry breaking \cite{Chernodub2}, coherence and quantum decoherence \cite{Huang,Wang}, quantum chromodynamics \cite{Yamamoto,Koepf}, pairing phase transitions \cite{Jiang}, carbon nanotubes \cite{Cunha2015}, fullerenes \cite{Lima2015}, etc. Also, for a more detailed discussion of the special relativity formalism in rotating frames, see Ref. \cite{Rizzi}.

In addition to noninertial effects, another type of effect that has also gained a lot of attention in the literature since the late 1970s are the topological effects generated by cosmic strings (CSs) \cite{Kibble,Vilenkin,Mello,Ade,Andrade,MSCunha,Cunha,Guvendi,Ahmed,Muniz}, where some effects are, for example, the quantization of energy \cite{Mazur}, self-forces \cite{Linet}, production of radiation by charged particles without acceleration \cite{Audretsch}, production of gravitational radiation (waves) \cite{Vilenkin1981,Vachaspati}, vacuum polarization \cite{Frolov}, gravitational lens \cite{Sazhin}, violation of causality via closed timelike curves (CTCs) \cite{Gott}, and the gravitational time delay \cite{Harari}. In particular, in addition to CSs being a ``special'' type of (hypothetical) linear gravitational topological defect, are highly dense, stable, infinitely long, and straight relativistic objects, can be static or spinning (with or without angular momentum), has cylindrical symmetry, is one of the exactly soluble models of Einstein's equations (solutions of a ``Kerr'' spacetime in (2+1)-dimensions), probably arose during the initial stages of the early universe (Big Bag) due to a cosmological phase transition with rotational symmetry-breaking, have a locally flat geometry (but not globally) with a conical singularity at the origin, and are mainly characterized by a planar angular deficit (the total angle around the CS is less than 360°) \cite{Kibble,Vilenkin,Mello,Ade,Andrade,MSCunha,Cunha,Guvendi,Ahmed,Muniz}. Besides, other interesting cosmological-like topological defects that could also have been formed in the early universe (via cosmological phase transitions with spontaneous symmetry-breaking) are domain walls (two-dimensional defects) \cite{Vilenkin}, magnetic monopoles (point defects) \cite{Vilenkin}, and cosmic vortons (loops of cosmic) \cite{Davis}. Now, from an observational point of view, there are some detectors that will try to look for CSs signals (gravitational waves emitted by oscillating CSs, for example \cite{Ruutu}), which are the Laser Interferometer Gravitational Wave Observatory (LIGO) \cite{Cui,Abbott}, the Laser Interferometer Space Antenna (LISA) \cite{Auclair,Blanco}, and the North American Nanohertz Observatory for Gravitational Waves (NANOGrav) \cite{Buchmuller,Blasi,Ellis}.

However, unlike cosmological-like topological defects, several topological defects in condensed matter (or condensed matter-like defects) have already been experimentally observed \cite{Kleinert,Ruutu,Katanaev}. In particular, one of these defects has some similarities with CSs, namely, disclinations, also called positive wedge disclinations, or simply conical defects (negative wedge disclinations do not have similarities with CSs) \cite{Vilenkin,Katanaev,Kleman,Kolesnikova}. So, in addition to CSs and disclinations being linear defects that change the topology of the medium in a similar way due to a rotational symmetry-breaking, the spatial part of the line element of a (static) CS corresponds to the line element of a disclination in liquid crystals (both have a planar angular deficit or positive curvature). In other words, a disclination is the nonrelativistic analogous (not the nonrelativistic limit) of a static CS for low energies. In that way, when we discuss the nonrelativistic limit of the DE, we can extend these discussions to the condensed matter physics context, where a CS could play the role of a disclination. By way of illustration, in Ref. \cite{Cortijo} the electronic properties of curved graphene in the static CS spacetime have already been studied through an effective DE (``relativistic DE''). Therefore, this shows that condensed matter physics can be an excellent laboratory for simulating various problems of gravitation and cosmology \cite{Ruutu,Zurek,Hendry,Moraes}.

On the other hand, in more recent years, some physicists have used the combination of the noninertial effects of rotating frames with the topological effects of CSs to study various systems of relativistic and nonrelativistic quantum mechanics. For example, from a relativistic point of view, this combination of the two effects has already been applied in the study of the Dirac and Klein-Gordon oscillators \cite{OliveiraGRG,Santos2018,Zare}, Aharonov-Casher effect \cite{Oliveira2019}, Aharonov-Bohm quantum rings \cite{GRG}, scalar bosons \cite{Castro}, geometric quantum phases \cite{BakkePRD}, etc. Now, from a nonrelativistic point of view, the combination of these two effects has already been applied in the study of quantum dots \cite{Rojas}, scattering \cite{Mota}, bound states for neutral particles with AMM and electric dipole moment \cite{Bakke2010}, and in the QHE \cite{Bran}. Recently, the Dirac, Klein-Gordon, and DKP-like oscillators have been studied under the influence of the noncommutativity and noninertial effects in a cosmic string spacetime \cite{Cuzinatto1,Cuzinatto2,Cuzinatto3}. However, in all these works, only the purely static case of the CS was considered, i.e., its intrinsic spin angular momentum or simply its angular momentum was not taken into account. Besides, in Refs. \cite{Cuzinatto1,Cuzinatto2,Cuzinatto3} the NC of the positions was also not taken into account. Therefore, for a more general description, in our work, we consider a CS with an angular momentum non-null and also the NC of the positions, respectively.

The present paper has as its goal to investigate the bound-state solutions of the noncommutative quantum Hall effect (NCQHE) with AMM in three different relativistic scenarios, namely: the Minkowski spacetime (inertial flat case), the spinning CS spacetime (inertial curved case), and the spinning CS spacetime with noninertial effects (noninertial curved case). Therefore, we work in two relativistic backgrounds, one from special relativity (Minkowski spacetime) and the other from general relativity (spinning CS spacetime). In particular, in the first two scenarios the frame of reference is inertial (we have the flat and curved inertial NCQHE), while in the third is noninertial, that is, there is a constant rotating frame in the spinning CS spacetime (we have the curved noninertial NCQHE). In this way, we can say that we are going to investigate the bound-state solutions under the influence of topological effects (or curvature), noninertial effects (or rotation), and NC effects (NC of positions and moments), respectively. However, although CS also has rotation (spin), here, the noninertial effects are generated exclusively by the rotating frame. With respect to bound-state solutions, we focus primarily on eigenfunctions (Dirac spinor and Schr\"{o}dinger wave function) and on energy eigenvalues (discrete energy spectrum or Landau levels), in other words, on solutions of a given (relativistic and nonrelativistic) eigenvalue equation. Besides, here we focused our study only on the case of the integer QHE in polar coordinates, and we used the noncommutative Dirac equation (NCDE) to model our three relativistic scenarios.

So, to include a rotating frame $S'$ in our problem, we apply a passive rotation on the angular coordinate as: $\varphi\to\varphi+\omega t$, where $\omega>0$ (rotation anticlockwise) is the constant angular velocity of the frame. Now, to include the spinning CS spacetime (fixing the background), we also made a change in the angular coordinate, and also in the time coordinate, that is: $\varphi\to\alpha\varphi$ and $t\to t+\beta\varphi$, where $\alpha\equiv 1-\frac{4G\bar{M}}{c^2}$ is a topological parameter (or curvature), $\beta\equiv\frac{4G\bar{J}}{c^3}$ is a rotational parameter (or rotation), where $\bar{M}>0$ is the linear mass density (mass per unit length), $\bar{J}>0$ the linear density of angular momentum (angular momentum per unit length) concentrated in the string core, and $G$ and $c$ are the gravitational constant and the speed of light, respectively 
\cite{Mazur,MSCunha,Cunha,Muniz}. Furthermore, to analytically solve our equations also for the third case, we consider two approximations: the first is that the linear velocity of the rotating frame is much less than the speed of light, and the second is that the coupling between the angular momentum of the CS and the angular velocity of the rotating frame is very weak (weak spin-rotation coupling for the CS).

The structure of this work is organized as follows. In Sect. \equ{sec2}, we make a brief review of the formalism of the NC phase spaces in $(2+1)$-dimensions (relativistic case) and in two-dimensions (nonrelativistic case). In Sect. \equ{sec3}, we introduce the NCDE in the Minkowski spacetime as well as explicitly obtain the relativistic and nonrelativistic bound-state solutions. In Sect. \equ{sec4}, we introduce the NCDE in a $(2+1)$-dimensional generic curved spacetime. In Sect. \equ{sec5}, we introduce the NCDE in the spinning CS spacetime as well as explicitly obtain the relativistic and nonrelativistic bound-state solutions. Already in Sect. \equ{sec5}, we introduce the NCDE in a rotating frame in the spinning CS spacetime as well as explicitly obtain the relativistic and nonrelativistic bound-state solutions (in the nonrelativistic case, we focus only on the spectrum). Finally, in Sect. \equ{sec5} we present our conclusions and some final remarks. In this work, we use the natural units where $\hbar=c=G=1$, and the spacetime with signature $(+,-,-)$.

\section{The noncommutative phase space}\label{sec2} 

In usual bidimensional (2D) usual quantum mechanics, a quantum phase space (or commutative phase space) is defined by substituting the classical canonical variables of position and momentum, given by $x_i$ and $p_j$, by their respective Hermitian operators, now written as $\hat{x}_i$ and $\hat{p}_j$, which obey the following Heisenberg (canonical) commutation relations (usual Heisenberg algebra) \cite{Szabo}
\begin{equation}\label{NC1}
[\hat{x}_i,\hat{p}_j]=i\delta_{ij}, \ \ [\hat{x}_i,\hat{x}_j]=0,\ \
[\hat{p}_i,\hat{p}_j]=0, \ \ (i,j=1,2=x,y),
\end{equation}
and whose Heisenberg uncertainty relations (Heisenberg uncertainty principle for position and momentum) are
\begin{equation}\label{NC2}
\Delta\hat{x}_i\Delta\hat{p}_j\geq\frac{1}{2}\delta_{ij}, \ \ \Delta\hat{x}_i\Delta\hat{x}_j=0, \ \ \Delta\hat{p}_i\Delta\hat{p}_j=0,
\end{equation}
where $\delta_{ij}=\delta_{ji}$ is the Kronecker delta (asymmetric tensor), also called Euclidean metric. Basically, the uncertainty principle states that we cannot measure simultaneously and with high precision two operators that do not commute with each other (incompatible operators). That is, the more we know about one operator (or observable), the less we know about the other (and vice versa).

Now, in order to define a noncommutative quantum phase space, or simply a NC phase space \cite{Bertolami,Berto,Bastos,Nascimento}, the relations in \eqref{NC1} must then obey the following deformed Heisenberg (noncanonical) commutation relations (NC or deformed Heisenberg algebra)
\begin{equation}\label{NC3}
[\hat{x}^{NC}_i,\hat{p}^{NC}_j]=i\delta_{ij}\left(1+\frac{\theta\eta}{4}\right), \ \ [\hat{x}^{NC}_i,\hat{x}^{NC}_j]=i\theta_{ij},\ \
[\hat{p}^{NC}_i,\hat{p}^{NC}_j]=i\eta_{ij},
\end{equation}
where the NC operators $\hat{x}^{NC}_i$ and $\hat{p}^{NC}_i$ are defined as
\begin{equation}\label{NC4}
\hat{x}^{NC}_i\equiv\hat{x}_i-\frac{1}{2}\theta_{ij}\hat{p}_j, \ \ \hat{p}^{NC}_i\equiv\hat{p}_i+\frac{1}{2}\eta_{ij}\hat{x}_j, \ \ (\hat{x}_i=\delta_{ij}\hat{x}^j; \ \hat{p}_j=\delta_{ij}\hat{p}^i=-i\partial_j),
\end{equation}
with $\theta_{ij}\equiv\theta\epsilon_{ij}$ and $\eta_{ij}\equiv\eta\epsilon_{ij}$ being antisymmetric constants ``tensors'' (real deformation parameters), $\epsilon_{ij}$ is the Levi-Civita symbol (a pseudotensor), and $\theta>0$ and $\eta>0$ are the position and momentum NC parameters with dimensions of length-squared and momentum-squared, respectively. From a phenomenological point of view, these two NC parameters can have the following values: $\theta\simeq 4.0 \times 10^{-40}$m$^2$ and $\eta\simeq 2.3 \times 10^{-61}$kg$^2$m$^2$s$^{-2}$ \cite{Berto}. In addition, it is worth mentioning here that an ``NC phase space'' can also arise naturally in condensed matter theory, i.e., in the QHE itself \cite{Szabo}. For example, how the total (linear) momentum for an electron ($q=-e$) in a constant magnetic field ($\vec{B}=B\vec{e}_z$) is given by: $\vec{\Pi}=\vec{p}-q\vec{A}$ (minimal coupling), where $\vec{p}=-i\vec{\nabla}$ is the canonical momentum and $\vec{A}=A_i\vec{e}_i$ is the vector potential with $A_i=-\frac{B}{2}\epsilon_{ij}x^j$ being its components, we have the following noncanonical commutation relations \cite{Szabo}
\begin{equation}\label{totalmomentum}
[\hat{x}_i,\hat{\Pi}_j]=i\delta_{ij}, \ \ [\hat{x}_i,\hat{x}_j]=0,\ \
[\hat{\Pi}_i,\hat{\Pi}_j]=iq\bar{B}\epsilon_{ij}, \ \ (i,j=1,2=x,y),
\end{equation}
where $\bar{B}=B$ is for the case of the symmetric gauge, given by: $\vec{A}=\frac{B}{2}(-y,x)$ \cite{Gusynin,Goldman,Dayi,Dulat}, and $\bar{B}=\frac{B}{2}$ is for the case of the Landau gauge, given by: $\vec{A}=\frac{B}{2}(-y,0)$ or $\vec{A}=\frac{B}{2}(0,x)$ (the choice is a matter of convenience) \cite{Haldane,Schakel,Beneventano,Lamata}. Therefore, as we see in (\ref{totalmomentum}), the momentum $\hat{\Pi}_{i(j)}$ does not commute with each other, consequently, the momentum space in the presence of a magnetic field becomes NC (we have a ``NC phase space''). 

In particular, the relationship between the set of NCs variables $\{\hat{x}^{NC}_i, \hat{p}^{NC}_i\}$ (here is not the anticommutator) with the set of usual variables $\{\hat{x}_i, \hat{p}_i\}$ is a consequence of a noncanonical linear transformation, known as Darboux transformation or Seiberg-Witten map \cite{Bertolami,Berto,Bastos,Nascimento}. However, all physical observables are entirely independent of the chosen map (how it should be). Furthermore, the NC space (NC only in position) causes a change in the usual product of two arbitrary functions $F(\vec{x})$ and $G(\vec{x})$, where now such a product is called of star product or Moyal product, and whose definition is given as follows \cite{Bertolami,Dayi}
\begin{equation}\label{NC5}
F(\vec{x},\theta)\star G(\vec{x},\theta)\equiv F(\vec{x})e^{(i/2)(\overleftarrow{\partial}_{x_i}\theta_{ij}\overrightarrow{\partial}_{x_j})}G(\vec{x})=F(\vec{x})e^{(i\theta/2)(\overleftarrow{\partial}_{x}\overrightarrow{\partial}_{y}-\overleftarrow{\partial}_{y}\overrightarrow{\partial}_{x})}G(\vec{x}),
\end{equation}
where it implies
\begin{equation}\label{NCv1}
[\hat{x}^{NC}_i,\hat{p}^{NC}_j]=\hat{x}^{NC}_i\star\hat{p}^{NC}_j-\hat{p}^{NC}_j\star\hat{p}^{NC}_i,
\end{equation}
\begin{equation}\label{NCv2}
[\hat{x}^{NC}_i,\hat{x}^{NC}_j]=\hat{x}^{NC}_i\star\hat{x}^{NC}_j-\hat{x}^{NC}_j\star\hat{x}^{NC}_i,
\end{equation}
\begin{equation}\label{NCv3}
[\hat{p}^{NC}_i,\hat{p}^{NC}_j]=\hat{p}^{NC}_i\star\hat{p}^{NC}_j-\hat{p}^{NC}_j\star\hat{p}^{NC}_i.
\end{equation}

In fact, in the absence of the NC space ($\theta_{ij}=0$), the star product is simply the usual product $F(x)\cdot G(x)$. Furthermore, the uncertainty relations for the NC case are now written as
\begin{equation}\label{NC6}
\Delta\hat{x}^{NC}_i\Delta\hat{p}^{NC}_j\geq\frac{1}{2}\delta_{ij}\left(1+\frac{\theta\eta}{4}\right), \ \ \Delta\hat{x}^{NC}_i\Delta\hat{x}^{NC}_j\geq\frac{1}{2}\vert\theta_{ij}\vert, \ \ \Delta\hat{p}^{NC}_i\Delta\hat{p}^{NC}_j\geq\frac{1}{2}\vert\eta_{ij}\vert.
\end{equation} 

It is interesting to mention that the first uncertainty relation in \eqref{NC6} lead to the appearance of an effective Planck constant which depends on the NC parameters $\theta$ and $\eta$, given in the form (restoring the factor $\hbar$)
\begin{equation}\label{NC7}
\hbar_{eff}=\hbar(1+\xi),
\end{equation}
where $\xi\equiv\frac{\theta\eta}{4\hbar^2}$. Then, for $\xi\ll 1$, or in the limit $\xi\to 0$, we recover the usual uncertainty relations. For a more detailed discussion of the possible hypothetical values of $\xi$, where was studied the NC gravitational quantum well, see Ref. \cite{Berto}.

Last but not least, the NC phase space can also be expanded to include the $(2+1)$-dimensional Minkowski spacetime of the relativistic quantum mechanics \cite{Berto}, this is
\begin{equation}\label{NC8}
[\hat{x}^{NC}_\mu,\hat{x}^{NC}_\nu]=i\left(\delta_{\mu\nu}+\frac{1}{4}\theta_{\mu}^{\sigma}\eta_{\nu\sigma}\right), \ \ [\hat{x}^{NC}_\mu,\hat{x}^{NC}_\nu]=i\theta_{\mu\nu},\ \
[\hat{p}^{NC}_\mu,\hat{p}^{NC}_\nu]=i\eta_{\mu\nu}, \ \ (\mu,\nu=0,1,2),
\end{equation}
where
\begin{equation}\label{NC9}
\hat{x}^{NC}_\mu=\hat{x}_\mu-\frac{1}{2}\theta_{\mu\nu}\hat{p}^\nu, \ \ \hat{p}^{NC}_\mu=\hat{p}_\mu+\frac{1}{2}\eta_{\mu\nu}\hat{x}^\nu, \ \ (\hat{x}_\mu=\tilde{\eta}_{\mu\nu}\hat{x}^\nu; \ \hat{p}_\mu=\tilde{\eta}_{\mu\nu}\hat{p}^\nu),
\end{equation}
and
\begin{equation}\label{NC10}
F(\vec{x},\theta)\star G(\vec{x},\theta)\equiv F(\vec{x})e^{(i/2)(\overleftarrow{\partial}_{x^{\mu}}\theta^{\mu\nu}\overrightarrow{\partial}_{x^{\nu}})}G(\vec{x}),
\end{equation}
being $\tilde{\eta}_{\mu\nu}=\tilde{\eta}^{\mu\nu}$= diag$(1,-1,-1)$ the flat metric tensor (Minkowski metric). Here, we use the indices given by the Greek letters to represent Minkowski spacetime (flat spacetime), as is usually done in the literature (in the absence of gravity). However, as in this work, we are also working on the curved spacetime; it is convention sometimes in the literature to use the indices given by the Greek letters ($\mu,\nu,\alpha,\beta,\ldots$) to represent a curved spacetime (this would be our general frame of reference) and the indices given by the Latin letters ($a, b, c, d,\ldots$) to represent a flat spacetime (this would be our local frame of reference). In addition, in this work, we consider the NC only in the spatial components of $\hat{x}^{NC}_{\mu}$ and $\hat{p}^{NC}_{\nu}$ (space-like NC), where it implies: $\theta_{0i}=\eta_{0j}=0$; otherwise, the unitarity (and causality or locality) of the quantum mechanics would not be preserved \cite{Berto,Szabo,Gomis}.

\section{The noncommutative Dirac equation in the Minkowski spacetime}\label{sec3}

The wave equation that governs the relativistic quantum dynamics of the QHE with AMM is given by the following tensorial DE with minimal and nonminimal couplings (corrected by QED and in Cartesian coordinates) \cite{Schakel,Geiger,Greiner,Greiner1}
\begin{equation}\label{M1}
\left[\gamma^a(p_a-qA_a)+\frac{\mu_{m}}{2}\sigma^{ab}F_{ab}-m_0\right]\Psi_D(t,\vec{r})=0, \ \ (a,b=t,x,y=0,1,2),
\end{equation}
where $\gamma^a=(\gamma^0,\vec{\gamma})$ are the flat Dirac gamma matrices and satisfies the anticommutation relation of the Clifford Algebra given by: $\{\gamma^a,\gamma^b\}=2\tilde{\eta}^{ab}I_{2\times 2}$, $p_a=i\partial_a=(p_0,-\vec{p})$ is the canonical momentum operator, $\sigma^{ab}=\frac{i}{2}[\gamma^a,\gamma^b]$ is an antisymmetric tensor, $F_{ab}=\partial_a A_b-\partial_b A_a$ is the electromagnetic field tensor, being $A_a=(A_0,-\vec{A})$ the electromagnetic potential (external electromagnetic field), $\Psi_D(t,\vec{r})\in\mathbb{C}^2$ is the two-component Dirac spinor (two-element column matrix), and $q<0$, $m_0$ and $\mu_{m}=\mu_{anomalous}\equiv a\mu_{B}$ are the electric charge (``minimal coupling constant''), rest mass, and the (anomalous) MDM of the fermion (``nonminimal coupling constant''), being $a$ the AMM and $\mu_B=\frac{e}{2m_0}>0$ is the famous Bohr magneton (MDM quantum), respectively. Furthermore, here we consider the most ``simple'' and elementary Dirac fermion (electron/positron), where the AMM is given by: $a_{e^-}=0.0011596521884$ and $a_{e^+}=0.0011596521879 $ \cite{Van}, and for simplicity we omit the symbol for quantum operators ( $\hat{•}$ ).

So, based on the fact that $\sigma^{ab}F_{ab}=i\gamma^i\gamma^j F_{ij}=-4\vec{S}\cdot\vec{B}$, where $\vec{B}=\vec{B}_{ext}$ is an external uniform magnetic field, $\vec{S}=\frac{1}{2}\vec{\Sigma}$ is the spin operator, with $\vec{\Sigma}=\vec{\sigma}$, being $\vec{\sigma}=(\sigma^1,\sigma^2,\sigma^3)$ the Pauli matrices (sigma matrices), and $F_{0i}=E_i=-\vert \nabla A_0\vert-\frac{\partial A_i}{\partial t}=0$ is the null electric field \cite{Greiner,Schluter,Villalba}, Eq. \eqref{M1} becomes
\begin{equation}\label{M2}
\left[\gamma^0 p_0+\gamma^i\left(p_i-q A_i\right)-2\mu_{m}\vec{S}\cdot\vec{B}-m_0\right]\Psi_D(t,\vec{r})=0, \ \ (i,j=1,2),
\end{equation}
or
\begin{equation}\label{M3}
\left[\gamma^0 p_0+\gamma^i\left(p_i+\frac{qB}{2}\epsilon_{ij}x^j\right)-2\mu_{m}\vec{S}\cdot\vec{B}-m_0\right]\Psi_D(t,\vec{r})=0,
\end{equation}
where $B=B_z=const.>0$ is the modulus or intensity of $\vec{B}$ and $A_i=-\frac{B}{2}\epsilon_{ij}x^j$ are the spatial components of the vector potential $\vec{A}$ \cite{Szabo,Bastos,Dayi,Dulat}.

To obtain the NCDE, it is necessary to write the operators $p_i$, $x^j$ and the spinor $\Psi_D$ in a NC phase space, which can be done as: $p_i\to p^ {NC}_i$, $x^j\to (x^j)^{NC}$ and $\Psi_D\to\star\Psi_D$. In this way, we have the following NCDE
\begin{equation}\label{M4}
\left[\gamma^0 p_0+\gamma^i\left(p^{NC}_i+\frac{qB}{2}\epsilon_{ij}(x^j)^{NC}\right)-2\mu_ {m}\vec{S}\cdot\vec{B}-m_0\right]\star\Psi_D(t,\vec{r})=0.
\end{equation}
or explicitly as ($\gamma^i p_i\to -\vec{\gamma}\cdot\vec{p};\ \gamma^i A_i\to -\vec{\gamma}\cdot\vec{A}$)
\begin{equation}\label{M5}
\left[\gamma^0 p_0-\vec{\gamma}\cdot(\tau\vec{p}+e\lambda\vec{A})-2\mu_{m}\vec{S}\cdot\vec{B}-m_0\right]\Psi^{NC}_D(t,\vec{r})=0,
\end{equation}
where we use
\begin{equation}\label{px}
p^{NC}_i=p_i+\frac{1}{2}\eta\epsilon_{im}x^m, \ \ (x^j)^{NC}=x^j-\frac{1}{2}\theta\epsilon^{jn}p_n,
\end{equation}
with $\tau$ and $\lambda$ being two real parameters defined as: $\tau\equiv\left(1-\frac{eB\theta}{4}\right)$ and $\lambda\equiv\left(1-\frac{\eta}{eB}\right)$, being $q=-e<0$ the charge of the electron, $\Psi^{NC}_D(t,\vec{r})$ is our NC Dirac spinor, and we also use the following relation of the Levi-Civita symbol in two spatial dimensions: $\epsilon_{ij}\epsilon^{mn}=\delta^m_i\delta^n_j-\delta^n_i\delta^m_j$, $\epsilon_{ij}\epsilon^{in}=\delta^n_j$ and $\epsilon_{ij}=-\epsilon_{ji}$ ($i,j,m,n=1,2$). However, for the second term in \eqref{M5} to not be null (physically impossible here), we must impose that: $\tau\neq 0$ (implies in $\theta\neq 4/m_0 \omega_c$ or $\omega_c\neq 4/m_0 \theta$) and $\lambda\neq 0$ (implies in $\eta\neq m_0 \omega_c$ or $\omega_c\neq\eta/m_0$). As we will see shortly (about the energy spectrum and degeneracy), these impositions will generate restrictions on the values of the magnetic field and also on the values of $\theta$ and $\eta$, in other words, only certain values are allowed depending of the values of the product $\tau\lambda$ ($\tau\lambda>0$ or $\tau\lambda<0$). In addition, the parameters $\tau$ and $\lambda$ are basically the same as those that appear in Ref. \cite{Dulat}, where was studied the nonrelativistic QHE in NC quantum mechanics under the influence of a uniform external electric field. 

On the other hand, in polar coordinates $(t,\rho,\varphi)$ where the line element takes the form: $ds^2_{Mink}=\tilde{\eta}_{ab}dx^{a} dx^{b}=dt^2-d\rho^2-\rho^ 2d\varphi^2$, being $\tilde{\eta}_{ab}=$diag$(1,-1,-\rho^2)$ ($a,b=t,\rho,\varphi$) the polar Minkowski metric, we have the following expressions for the momentum operator $\vec{p}$ and the vector potential $\vec{A}$ written in this curvilinear coordinates system
\begin{equation}\label{M6}
\vec{p}=-i\left(\vec{e}_\rho \partial_\rho+\frac{\vec{e}_\varphi}{\rho}\partial_\varphi\right), \ \ \vec{A}=\frac{1}{2}\vec{B}\times\vec{r}=A_\varphi\vec{e}_\varphi,
\end{equation}
where $-\infty<t<\infty$ is the time coordinate, $0\leq\varphi\leq 2\pi$ is the angular coordinate and $\rho=\sqrt{x^2+y^2}$, with $0\leq\rho<\infty$, is the polar radial coordinate, respectively.

Thus, substituting \eqref{M6} in (\ref{M5}), we get the following NCDE in polar coordinates
\begin{equation}\label{M7}
\left[i\gamma^0\partial_t+i\tau\gamma^\rho\partial_\rho+\gamma^\varphi\left(\frac{i\tau}{\rho}\partial_\varphi-e\lambda A_\varphi\right)-2\mu_{m}\vec{S}\cdot\vec{B}-m_0\right]\Psi^{NC}_D(t,\rho,\varphi)=0,
\end{equation}
where the curvilinear gamma matrices are defined in the form
\begin{equation}\label{M8}
\gamma^\rho (\varphi)=\vec{\gamma}\cdot\vec{e}_\rho=\gamma^1\cos\varphi+\gamma^2\sin\varphi, \ \ \gamma^\varphi(\varphi)=\vec{\gamma}\cdot\vec{e}_\varphi=-\gamma^1\sin\varphi+\gamma^2\cos\varphi,
\end{equation}
and satisfying the anticommutation relation of a ``curvilinear Clifford Algebra'', given by: $\{\gamma^{a},\gamma^{b}\}=2 \tilde{\eta}^{ab}I_{2\times 2}$.

However, due to the presence of $\cos\varphi$ and $\sin\varphi$, it is difficult to proceed without a simplification of Eq. (\ref{M7}). To eliminate this obstacle, it is necessary to use a similarity transformation (rotate the spinor), given by a unitary operator
$U(\varphi)=e^{-\frac{i\varphi\Sigma^3}{2}}\in SU(2)$ ($\Sigma^3=\Sigma_3=i\gamma^1\gamma^2$) \cite{Schluter,Villalba,Villalba2003}. As a consequence, we can convert through this operator the curvilinear gamma matrices $\gamma^\rho$ and $\gamma^\varphi$ into the fixed Cartesian gamma matrices $\gamma^1$ and $\gamma^2$ as follows \cite{Schluter,Villalba,Villalba2003}
\begin{equation}\label{transformacao}
U^{-1}\gamma^\rho U=\gamma^1, \ \ U^{-1}\gamma^\varphi U=\gamma^2.
\end{equation}

Therefore, using the information above and the fact that in the QHE the angular component of the vector potential associated with a constant uniform magnetic field $\vec{B}=B\vec{e}_z$ is given by $A_\varphi(\rho)=\frac{1}{2}B\rho$ (symmetric gauge in polar coordinates) \cite{Filgueiras,Brand,Villalba,Villalba2003}, we can then rewrite Eq. \eqref{M6} in the form
\begin{equation}\label{M9}
\left[i\gamma^0\partial_t+i\tau\gamma^1\left(\partial_\rho+\frac{1}{2\rho}\right)+\gamma^2\left(\frac{i\tau}{\rho}\partial_\varphi-\frac{\lambda eB}{2}\rho\right)-\mu_{m}\Sigma^3 B-m_0\right]\psi^{NC}_C(t,\rho,\varphi)=0,
\end{equation}
is our NC curvilinear Dirac spinor (transformed or rotated spinor), and both spinors must satisfy the following conditions: $\psi^{NC}_C(\varphi\pm 2\pi)=-\psi^{NC}_C(\varphi)$ e $\Psi^{NC}_D(\varphi\pm 2\pi)=\Psi^{NC}_D(\varphi)$. In particular, these conditions show that the original Dirac spinor is a continuous (everywhere) and periodic function whose period is $\pm 2\pi$ (final state = initial state), and the curvilinear spinor is only a periodic continuous function if the period is $\pm 4\pi$, that is: $\psi^{NC}_C(\varphi\pm 4\pi)=\psi^{NC}_C(\varphi)$ (final state = initial state).

Since we are working in a $(2+1)$-dimensional spacetime, it is need to define the gamma matrices $\gamma^a=(\gamma^0,\gamma^1,\gamma^2)=(\gamma_0,-\gamma_1,-\gamma_2)$ and the matrix $\Sigma^z=\Sigma^3$ directly in terms of the Pauli matrices $2\times 2$, in other words, we must have $\gamma_1=\sigma_3\sigma_1= i\sigma_2$, $\gamma_2=s\sigma_3\sigma_2=-is\sigma_1$ and $\gamma^0=\Sigma^3=\sigma_3$, where $s=\pm 1$ is a parameter called the spin parameter and describes the two spin states of the planar fermion: $s=+1$ is for the spin ``up'' (``$\uparrow$'') and $s=-1$ is for the ``down'' spin (``$\downarrow$''), respectively \cite{Villalba,Villalba2003,Hagen,Oliveira2019,OliveiraGRG,Andrade}. Here, it is important to note that $s$ (an ``unfortunate'' notation by the way) is not technically the spin quantum number $s=1/2$, or the spin magnetic quantum number $m_s=\pm 1/2= \uparrow\downarrow$. However, we can easily write $s$ in terms of $m_s$ (and vice versa), that is: $s\equiv 2m_s$ (or $m_s\equiv s/2$). Furthermore, this spin parameter came up in order to show that there is an exact equivalence between the Aharonov-Bohm effect for spin-1/2 particles and the Aharonov-Casher effect. So, as both phenomena are purely planar, such equivalence would only be possible if only the vector potential and the electric field were dual to each other, and therefore, the following relation must be satisfied: $eA_i=s\mu_m\epsilon_{ij}E^j$ ($\mu_m=a\mu_{B}$), where $A_i$ are the components of the vector potential of the Aharonov-Bohm effect and $E^j$ are the components of the electric field of the Aharonov-Casher effect, respectively \cite{Hagen}.

Now, defining the following (stationary) ansatz for the two-component curvilinear spinor \cite{Greiner,Schluter,Villalba,Villalba2003}
\begin{equation}\label{spinor}
\psi^{NC}_C(t,\rho,\varphi)=\frac{e^{i(m_j\varphi-Et)}}{\sqrt{2\pi}}\left(
           \begin{array}{c}
            f(\rho) \\
            ig(\rho) \\
           \end{array}
         \right),
\end{equation}
and knowing that the Pauli matrices take the form
\begin{equation}\label{matrices}
\sigma_1=\sigma_x=\left(
    \begin{array}{cc}
      0\ &  1 \\
      1\ & 0 \\
    \end{array}
  \right), \ \  \sigma_2=\sigma_y=\left(
    \begin{array}{cc}
      0 & -i  \\
      i & \ 0 \\
    \end{array}
  \right), \ \  \sigma_3=\sigma_z=\left(
    \begin{array}{cc}
      1 & \ 0 \\
      0 & -1 \\
    \end{array}
  \right),
\end{equation}
we obtain from (\ref{M9}) a set of two coupled differential equations of the first order, given by
\begin{equation}\label{M10}
\frac{(m_0+\mu_m B-E)}{\tau}f(\rho)=\left[\frac{d}{d\rho}+sm_0\Omega\rho+\frac{s}{\rho}\left(m_j+\frac{s}{2}\right)\right]g(\rho),
\end{equation}
\begin{equation}\label{M11}
\frac{(m_0-\mu_m B+E)}{\tau}g(\rho)=\left[\frac{d}{d\rho}-sm_0\Omega\rho-\frac{s}{\rho}\left(m_j-\frac{s}{2}\right)\right]f(\rho),
\end{equation}
where $f(\rho)$ and $g(\rho)$ are real radial functions, $\Omega=\Omega_{eff}\equiv\frac{\lambda\omega_c}{2\tau}$ is a effective angular frequency, with $\tau=(1-m_0\omega_c\theta/4)$ and $\lambda=(1-\eta/m_0\omega_c)$, being $\omega_c=\frac{eB}{m_0}>0$ the famous cyclotron frequency (an angular velocity), $E$ is the relativistic total energy of the fermion (or antifermion), $\frac{1}{\sqrt {2\pi}}$ is a constant factor that arises from normalizing of the angular part of the spinor, and $m_j=\pm\frac{1}{2},\pm\frac{3}{2},\pm\frac {5}{2},\ldots$ is the total magnetic quantum number, which arises from the ``periodicity condition'' of the curvilinear spinor, that is, $ \psi^{NC}_C(\varphi\pm 2\pi)=-\psi^{NC}_C(\varphi)$ \cite{Schluter,Villalba}. In addition, the connection between $m_j$ and $m_l$, where $m_l$ is the orbital magnetic quantum number, is given by
\begin{equation}\label{M12}
J_z\psi^{NC}_C(t,\rho,\varphi)=-i\partial_\varphi\psi^{NC}_C(t,\rho,\varphi)=m_j\psi^{NC}_C(t,\rho,\varphi)=(m_l+m_s)\psi^{NC}_C(t,\rho,\varphi),
\end{equation}
where $J_z=L_z+S_z$, $m_s=\pm\frac{1}{2}\equiv\frac{s}{2}$ and the values of $m_l$ are $m_l=0,\pm 1, \pm 2,\ldots$. In particular, a consequence of the values $m_j$ and $m_l$ is that $L_z$ and $S_z$ can be parallel or antiparallel (excluding $m_l=0$), i.e., if in $J_z$ we have $m_l>0 $ and $m_s=+1/2$ or $m_l<0$ and $m_s=-1/2$, then $L_z$ and $S_z$ are parallel. Now, if in $J_z$ we have $m_l>0$ and $m_s=-1/2$ or $m_l<0$ and $m_s=+1/2$, then $L_z$ and $S_z$ are antiparallel. As we will see shortly, this will directly influence the energy spectrum as well as in its degeneracy.

Therefore, substituting (\ref{M11}) into (\ref{M10}), we get the following second-order differential DE (``quadratic DE'') for the NCQHE with AMM in the Minkowski spacetime
\begin{equation}\label{M13}
\left[\frac{d^2}{d\rho^2}+\frac{1}{\rho}\frac{d}{d\rho}-\frac{\Gamma^2}{\rho^2}-(m_0\Omega\rho)^2+\mathcal{E}\right]f(\rho)=0,
\end{equation}
where we define
\begin{equation}\label{M14}
\Gamma\equiv\left(m_j-\frac{s}{2}\right), \ \ \mathcal{E}\equiv\frac{(E-\mu_m B)^2-m_0^2}{\tau^2}-2m_0\Omega\left(m_j+\frac{s}{2}\right).
\end{equation}

\subsection{Bound-state solutions: two-component Dirac spinor and relativistic Landau levels}\label{subsec1}

To analytically solve Eq. \eqref{M13}, let's introduce a new (dimensionless) variable in our problem, given by: $r=m_0\vert\Omega\vert\rho^2>0$. In that way, Eq. \eqref{M13} becomes
\begin{equation}\label{M15}
\left[r\frac{d^{2}}{dr^{2}}+\frac{d}{dr}-\frac{\Gamma^{2}}{4r}-\frac{r}{4}+\frac{\mathcal{E}}{4m_0\vert\Omega\vert}\right]f(r)=0.
\end{equation}

Now, analyzing the asymptotic behavior (or asymptotic limit) of Eq. (\ref{M15}) for $r\to 0$ and $r\to\infty$, we get a (regular) solution to this equation given by the following ansatz
\begin{equation}\label{M16}
f(w)=Cr^{\frac{\vert\Gamma\vert}{2}}e^{-\frac{r}{2}}F(r), \ \ (\vert\Gamma\vert\geq 0),
\end{equation}
where $C>0$ is a normalization constant, $F(r)$ is an unknown function to be determined, and $f(r)$ must satisfy the following boundary conditions to be a normalizable solution (finite solution)
\begin{equation}\label{conditions} 
f(r\to 0)=f(r\to\infty)=0.
\end{equation}

Substituting \eqref{M16} in \eqref{M15}, we have a differential equation for $F(r)$ as follows
\begin{equation}\label{M17}
\left[r\frac{d^{2}}{dr^{2}}+(\vert\bar{\Gamma}\vert-r)\frac{d}{dr}-\mathcal{\bar{E}}\right]F(r)=0,
\end{equation}
where
\begin{equation}\label{M18}
\vert\bar{\Gamma}\vert\equiv\vert\Gamma\vert+1, \ \ \mathcal{\bar{E}}\equiv\frac{\vert\bar{\Gamma}\vert}{2}-\frac{\mathcal{E}}{4m_0\vert\Omega\vert}.
\end{equation}

Before proceeding, let's make a small observation about one of the boundary conditions. For example, if in $w$ we had used $\Omega$ instead of $\vert\Omega\vert$, then it would imply in $f(w\to\infty)\to\infty$ with $\Omega<0$ (here this frequency can be negative). Consequently, we would have a solution not normalizable, something impossible because we are looking for bound-state solutions. So, according to the literature \cite{Villalba}, Eq. \eqref{M17} is an (generalized) associated Laguerre equation, whose solution are the (generalized) associated Laguerre polynomials $F(r)=L^{\vert\Gamma\vert}_n(r)$. Consequently, the quantity $\mathcal{\bar{E}}$ must to be equal to a non-positive integer, i.e., $\mathcal{\bar{E}}=-n$ (quantization condition), where $n=n_\rho=0,1,2,\ldots$ is a (radial) quantum number. Therefore, from this condition, we obtain the following relativistic energy spectrum (relativistic Landau levels) for the NCQHE with AMM in the Minkowski spacetime
\begin{equation}\label{spectrum}
E^{\kappa}_{n,m_j,s,s'}=E_m+\kappa\sqrt{m_0^2+2m_0\vert\tau\lambda\vert\omega_c N},
\end{equation}
with
\begin{equation}\label{M19}
E_m=E^{\kappa}_m\equiv\mu_mB>0, \ \	N=N_{eff}\equiv\left(n+\frac{1}{2}+\frac{\big|m_j-\frac{s}{2}\big|+s'\left(m_j+\frac{s}{2}\right)}{2}\right)\geq 0,
\end{equation}
and
\begin{equation}\label{taulambda}
\tau\lambda=\frac{1}{4m_0\omega_c}(4-m_0\omega_c\theta)(m_0\omega_c-\eta),
\end{equation}
where $\kappa=\pm 1$ is a parameter (``energy parameter'') that represents the positive energy states ($\kappa=+1$) or particle states (electron), as well as the negative energy states ($\kappa=-1$) or antiparticle states (positron), $N$ is an effective quantum number (as it depends on all others), and we using $\Omega=s'\vert\Omega\vert$, where $s'$=sign$(\tau\lambda)=\pm 1$ is a parameter that describes the positive and negative signs (values) of the product $\tau\lambda$ ($\tau\lambda>0$ or $\tau\lambda<0$). So, we see that in addition to spectrum \equ{spectrum} being asymmetric, i.e., the energies of the particle and antiparticle are not equal ($E^+\neq\vert E^-\vert$), such spectrum depends linearly on the magnetic (potential) energy $E_m$ generated by the interaction of the AMM (or MDM) with the external magnetic field $B$ (is the cause of the ``break'' of symmetry of the spectra), where such energy has approximately the same value (up to 11 decimal places) for the particle and antiparticle ($E^+_m\simeq E^-_m$), as well as explicitly depends on the quantum numbers $n$ and $m_j$, cyclotron frequency $\omega_c$, spin parameter $s$, and on the NC parameters $\theta$ and $\eta$, respectively. It is interesting to note that the quantization of energy arises from the so-called Landau quantization, which refers to the quantization of the ``cyclotronic orbits'' of charged particles in magnetic fields, where such particles can only occupy ``orbits'' with discrete values of energy, called Landau levels. Besides, we note that even in the absence of the magnetic field ($B=0$), the spectrum still remains quantized (discrete) due to the presence of the parameter $\eta$, which acts as a kind of ``NC field'', and is given by: $E^{\kappa}_{n,m_j,s}=\kappa\sqrt{m_0^2+2\eta N}$ (does not depend on $s'$).

Before analyzing graphically and in detail the behavior of the spectrum \eqref{spectrum} as a function of the magnetic field $B$ and of the NC parameters $\theta$ and $\eta$ for different values of $n$ and $m_j$, it is advisable to first analyze one of the most important aspects of the two-dimensional energy spectra (or three-dimensional), which is its degeneracy or the degenerate states. In particular, we verify that this degeneracy depends on the values (signs) of $s'$, $s$, and $m_j$. In other words, by fixing a given value of $s'$ in $N$, we can know the degeneracy according to the values of $s$ and $m_j$. In addition, it is important to comment that fixing a given $m_j$ ($m_j>0$ or $m_j<0$), we can have two spin states ($s=+1$ or $s=-1$). Therefore, in Table \ref{tab1} we have eight possible configurations for the degeneracy depending on the values of $s'$, $s$, and $m_j$, as well as the respective values of $N$ and $m_l$ and also the orientation of the operators $L_z$ and $S_z$.
\begin{center}
\begin{table}[h]
\centering
\caption{Degeneracy depends on the values of $s'$, $s$ and $m_j$.}
\def\arraystretch{1.1}
\begin{tabular}{cccccccc}
\hline
Configuration &$s'$ & $m_j$ & $s$ & $N$ & $m_l$ & Degeneracy & Orientation of $L_z$ and $S_z$ \\ \hline
1& +1 & $m_j>0$ & $+1$ & $n+m_l+1$ & $m_l\geq 0$& finite & parallel \\
2& +1 & $m_j>0$ & $-1$ & $n+m_l$ & $m_l\geq 1$& finite& antiparallel  \\
3& +1 & $m_j<0$ & $+1$ & $n+1$ & $m_l<0$& infinite& antiparallel  \\
4& +1 & $m_j<0$ & $-1$ & $n$ & $m_l\leq 0$& infinite& parallel  \\
5& $-1$ & $m_j>0$ & $+1$ & $n$ & $m_l\geq 0$& infinite& parallel  \\
6& $-1$ & $m_j>0$ & $-1$ & $n+1$ & $m_l\geq 1$& infinite& antiparallel  \\
7& $-1$ & $m_j<0$ & $+1$ & $n+\vert m_l\vert$ & $m_l<0$& finite& antiparallel  \\
8& $-1$ & $m_j<0$ & $-1$ & $n+\vert m_l\vert+1$ & $m_l\leq 0$& finite& parallel 
\\ \hline
\end{tabular}
\label{tab1}
\end{table}
\end{center}

According to Table \ref{tab1}, we see that the spectrum can be finitely or infinitely degenerate, and therefore, there can be a finite or infinite number of degenerate states (states sharing the same energy) depending on the values of $s'$, $s$, and $m_j$, respectively. In particular, infinite degeneracy arises when the spectrum only depends on the quantum number $n$, and since $m_l$ can take any integer value in $m_j=m_l+s/2$, it implies that the states are infinitely degenerate \cite{Rubens}. With respect to finite degeneracy, such degeneracy arises when the spectrum depends on both the quantum numbers $n$ and $m_j$ (or $m_l$), where it is now possible to define a new quantum number from $n$ and $m_l$, given by: $k\equiv n+m_l\geq 1$ ($m_l\geq1$), or $k\equiv n+m_l\geq 0$ ($m_l\geq0$). Then, from the number $k$ we can determine the expression for the total degree of degeneracy for each energy level $E_k$, which is given by: $\Omega(k)=\sum\limits_{m_l=1}^{k}(2m_l+1)=k(k+2)$, or $\Omega(k)=\sum\limits_{m_l=0}^{k}(2m_l+1)=(k+1)^2$, where $2m_l+1$ is the (finite) number of degenerate states of the system \cite{Rubens}. Thus, we see that the ground state ($n=0$) for $m_l=0$ ($k=0$ and $\Omega(0)=1$) is non-degenerate, while for $m_l=1$ ($k=1$ and $\Omega(1)=3$) is triple degenerate. In this way, we explicitly see that degeneracy plays a key role in the spectra of the NCQHE, and possibly even in its thermodynamic properties since we can build a canonical ensemble for the system \cite{Rubens}.

So, based on the informations about the degeneracy (Table \ref{tab1}) we get Table \ref{tab2}, where we have eight possible configurations (eight for both particle and antiparticle) for the spectrum depending on the values of $s'$, $s$, and $m_j$.
\begin{center}
\begin{table}[h]
\centering
\caption{Energy spectra for the degenerate states of the particle and antiparticle.}
\def\arraystretch{1.1}
\begin{tabular}{ccc}
\hline
Configuration & Energy spectrum $E^{\kappa}_{n,m_j,s,s'}$ & Degeneracy \\ \hline
1 &$E^{\kappa}_{n,m_j>0,+,+}=E_m+\kappa\sqrt{m_0^2+2m_0\vert\tau\lambda\vert\omega_c(n+m_l+1)}$ & finite \\
2 &$E^{\kappa}_{n,m_j>0,-,+}=E_m+\kappa\sqrt{m_0^2+2m_0\vert\tau\lambda\vert\omega_c(n+m_l)}$ & finite \\
3 &$E^{\kappa}_{n,m_j<0,+,+}=E_m+\kappa\sqrt{m_0^2+2m_0\vert\tau\lambda\vert\omega_c(n+1)}$ & infinite \\
4 &$E^{\kappa}_{n,m_j<0,-,+}=E_m+\kappa\sqrt{m_0^2+2m_0\vert\tau\lambda\vert\omega_c n}$ & infinite \\
5 &$E^{\kappa}_{n,m_j>0,+,-}=E_m+\kappa\sqrt{m_0^2+2m_0\vert\tau\lambda\vert\omega_c n}$ & infinite \\
6 &$E^{\kappa}_{n,m_j>0,-,-}=E_m+\kappa\sqrt{m_0^2+2m_0\vert\tau\lambda\vert\omega_c (n+1)}$ & infinite \\
7 &$E^{\kappa}_{n,m_j<0,+,-}=E_m+\kappa\sqrt{m_0^2+2m_0\vert\tau\lambda\vert\omega_c(n+\vert m_l\vert)}$ & finite \\
8 & $E^{\kappa}_{n,m_j<0,-,-}=E_m+\kappa\sqrt{m_0^2+2m_0\vert\tau\lambda\vert\omega_c(n+\vert m_l\vert+1)}$ & finite
\\ \hline
\end{tabular}
\label{tab2}
\end{table}
\end{center}

According to Table \ref{tab2}, we see that for $s'=+1$ with $m_j>0$, or $s'=-1$ with $m_j<0$ (configs. 1, 2, 7, and 8), the spectra have exactly the same energy values, regardless of the spin chosen, consequently, the spin does not change the values of the spectra. Furthermore, these are the only cases where the spectra depend on both $n$ and $m_j$, and therefore, the energies increase as a function of $n$ and $m_j$. However, for $s'=+1$ with $m_j<0$, or $s'=-1$ with $m_j>0$, then the spectrum is already influenced by the spin, that is, depending on the spin chosen the energies can be higher or lower (configs. 3-6). For example, the energies of the particle ($\kappa=+1$) are greater when its spin is up (config. 3), while for the antiparticle ($\kappa=-1$) the energies are greater (in absolute values) when its spin is down (config. 6). Furthermore, regardless of the config. chosen, the energies of the particle are always higher than those of the antiparticle. In particular, this is due to the presence of the magnetic energy $E_m$, in which ``breaks'' the symmetry of the spectra, i.e., the spectra of the particle and antiparticle do not have the same values. Now, we are ready to compare the spectrum (\ref{spectrum}) with the literature, or rather, the spectra from Table \ref{tab2}. So, we verified that in the absence of the NC phase space ($\theta=\eta=0$) and of the AMM ($E_m=0$), with the quantum number now being $n'=n+1=1,2,\ldots,\infty$, we obtain exactly the (symmetric) usual spectrum of the relativistic QHE for $m_0\neq 0$ (massive case) \cite{Haldane,Schakel,Lamata} or for $m_0=0$ (massless case) \cite{Beneventano}, or even the ``relativistic'' spectrum of the QHE for the graphene \cite{Novoselov,Gusynin,Goldman,Katsnelson}, in which we should replace: $c\to v_f$ and $m_0\to\Delta\geq 0$, where $v_f\sim\frac{c}{300}$ is the Fermi velocity and $\Delta$ is a mass gap (or ``effective mass''). From the above, we clearly see that our relativistic spectrum generalizes many particular cases of the literature.

On the other hand, even with the well-defined spectra in Table \ref{tab2} we are not yet ready to analyze the behavior of the relativistic spectrum as a function of $B$, $\theta$ and of $\eta$ for different values of $n$, i.e., we first need to define the restrictions for these three quantities. As discussed in the previous section, the parameters $\tau$ and $\lambda$ must obey the following conditions (restrictions): $\tau\neq 0$, where implies that $\theta\neq 4/m_0 \omega_c$ or $\omega_c\neq 4/m_0 \theta$, and $\lambda\neq 0$, where implies that $\eta\neq m_0 \omega_c$ or $\omega_c\neq\eta/m_0$, consequently, this implies that the product $\tau\lambda$ can be positive ($\tau\lambda>0$ or $s'=+1$) or negative ($\tau\lambda<0$ or $s'=-1$), respectively. In that way, by restricting the NC parameters ($\theta\neq 4/m_0 \omega_c$ and $\eta\neq m_0 \omega_c$), the field is arbitrary, while by restricting the field ($\omega_c\neq 4/m_0 \theta$ and $\omega_c\neq\eta/m_0$), are the NC parameters that are arbitrary. Then, using the expression of the product $\tau\lambda$, given by \eqref{taulambda} (explicitly is a quadratic inequality for $B$), we get Table \ref{tab3}, where we have two possible restrictions for $B$ depending of the values of $s'$.
\begin{center}
\begin{table}[h]
\centering
\caption{Restrictions for the magnetic field $B$.}
\def\arraystretch{1.1}
\begin{tabular}{cccc}
\hline
Restrictions & $s'$ & $B$ \\ \hline
1 & $+1$ & $\frac{(4+\theta\eta)-\sqrt{(4+\theta\eta)^2-16\theta\eta}}{2e\theta}<B<\frac{(4+\theta\eta)+\sqrt{(4+\theta\eta)^2-16\theta\eta}}{2e\theta}$ \\
2 & $-1$ & $\frac{(4+\theta\eta)+\sqrt{(4+\theta\eta)^2-16\theta\eta}}{2e\theta}<B<\frac{(4+\theta\eta)-\sqrt{(4+\theta\eta)^2-16\theta\eta}}{2e\theta}$  \\ \hline
\end{tabular}
\label{tab3}
\end{table}
\end{center}

According to Table \ref{tab3}, we can analyze the behavior of the spectrum as a function of $B$ for each one of the two restrictions, or for convenience (and without loss of generality), only one restriction. In particular, these ranges define the regions allowed for the NCQHE to manifest (to exist), where the regions outside these intervals are forbidden regions. In addition, if in our problem we only had the NC of the position ($\eta=0$), then the range for $B$ would be only $0<B<4/e\theta$ (for $s'=+1$), while the range $4/e\theta<B<0$ (for $s'=-1$) would be discarded because here we do not admit negative fields . Now, in the absence of the NC phase space ($\theta=\eta\to 0$), then the range for $B$ is given by $0\leq B<\infty$, and therefore, is a physically consistent result since in the usual QHE the field can vary between zero and ``infinity''. In this way, we see that the two ranges of Table \ref{tab3} do not generate any physical inconsistency, since in the presence of the NC phase space we have $(4+\theta\eta)^2-16\theta\eta>0$ (real magnetic fields), and in the absence we recover the usual range of $B$ ($0\leq B<\infty$).  

\begin{figure}[ht]
\centering
\includegraphics[scale=1.0]{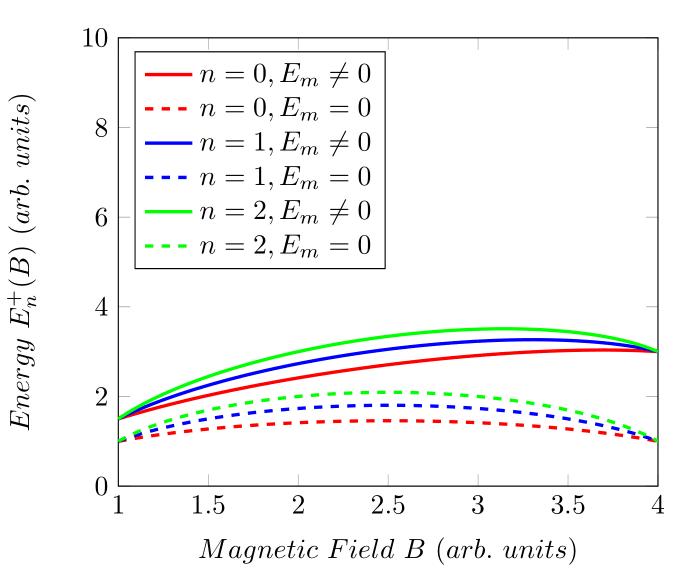}
\caption{Graph of $E^+_n(B)$ versus $B$ for three different values of $n$ with $E_m\neq 0$ ($a=1$) and $E_m=0$ ($a=0$).}
\label{fig1}
\end{figure}

Therefore, now we can analyze in detail the behavior of the spectrum as a function of the magnetic field for different values of $n$ (we omit $m_l$ here because its function is analogous to that of $n$). For the sake of practicality and simplicity, we do not need to analyze the behavior of all the spectra of Table (\ref{tab2}) and much less use the two ranges of Table (\ref{tab3}); only one spectrum for the particle and antiparticle and one range is enough for our purpose. In this way, we can focus on the ``simple'' spectrum where the ground state ($n=0$) still depends on $\omega_c$, $\theta$, and $\eta$, that is, the spectra of configs. 3 or 6. Therefore, choosing config. 3 with restriction 1 (also used for all graphs here), we have Fig. \ref{fig1}, where shows the behavior of the energies of the particle as a function of the magnetic field for the ground state ($n=0$) and the first two excited states ($n=1,2$), with and without the presence of magnetic energy $E_m$, in which we take for simplicity that $m_0=e=\theta=\eta=1$, where the allowed range for the field is $1<B<4$. 

According to \ref{fig1}, we see that the energies increase with the increase of $n$ (as it should be), and are greater in the presence of magnetic energy. Therefore, the function of the AMM is to increase the energies of the particle. Also, the energies can increase or decrease as a function of $B$. For example, for the case $E_m\neq 0$ the energies increase between $B\approx 1$ and $B\approx 3.3$ (but with the exception of $n=0$), and decrease between $B\approx 3.3$ and $B\approx 4$ (with $\Delta E=E_{final}-E_{initial}>0$), while for the case $E_m=0$ the energies increase between $B\approx 1$ and $B=2.5$, and decrease between $B=2.5$ and $B\approx 4$ (with $\Delta E=E_{final}-E_{initial}=0$).  

\begin{figure}[ht]
\centering
\includegraphics[scale=1.0]{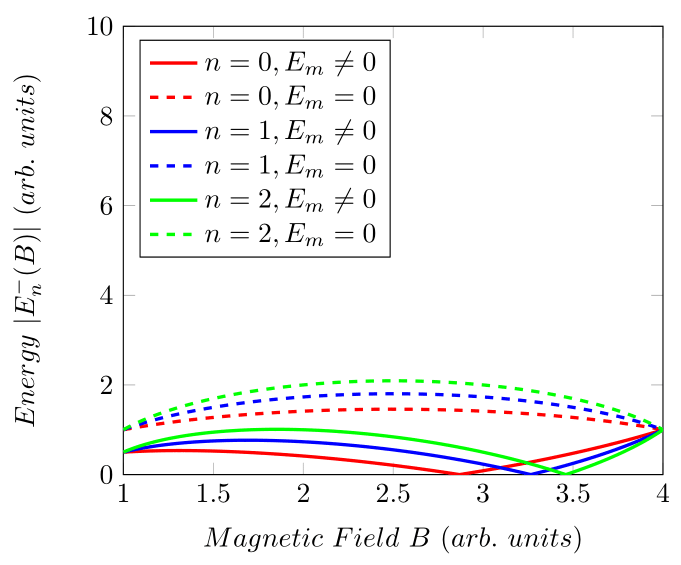}
\caption{Graph of $\vert E^{-}_n (B)\vert$ versus $B$ for three different values of $n$ with $E_m\neq 0$ ($a=1$) and $E_m=0$ ($a=0$).}
\label{fig2}
\end{figure}

In Fig. \ref{fig2}, we see the behavior of the energies of the antiparticle as a function of the magnetic field for the ground state and the first two excited states (with $m_0=e=\theta=\eta=1$ and $1<B<4$). According to this Figure, we see that the energies for the case $E_m=0$ are basically equal to those of the particle also with $E_m=0$ (symmetric spectra), and therefore, the energies increase with the increase of $n$, and $E_{final}=E_{initial}$ ($\Delta E=0$). However, these energies are larger in the absence of magnetic energy, i.e., the function of the AMM is to decrease the energies of the antiparticle. Now, for the case $E_m\neq 0$, the energies can increase or decrease with the increase of $n$. For example, between $B\approx 1$ and $B\approx 3.1$, the energies are higher with the increase of $n$. In particular, the approximate value of $B$ ($B\approx 3.1$) was calculated through the crossing of the red and blue solid lines, where their respective energies are equal (equal spectra). Now, between $B\approx 3.4$ and $B\approx 4$, the energies are smaller with the increase of $n$ (an ``anomaly''). Again, the value of $B\approx 3.4$ was calculated through the crossing of the green and blue solid lines. Moreover, the energies can increase, decrease, or be ``null'' (``zero'') as a function of $B$ (case $E_m\neq 0$). Here, ``null energy'' does not mean that the antiparticle has no energy; it means that the resulting total energy is null, i.e., for certain values of $B$, the magnetic energy is equal to the quantized energy (``square root energy''). In particular, such ``null energies'' appear when the solid lines touch the axis-$B$ and then increase until $B\approx 4$, where $E_{final}>E_{initial}$ ($\Delta E>0$). Then, solving $\vert E^{-}_{n}(B)\vert=0$ for each specific state, we have $B\approx 2.9$ ($n=0$), $B\approx 3.3$ ($n=1$), and $B\approx 3.4$ ($n=2$), respectively.

Now, let's analyze the behavior of the spectrum as a function of the NC parameters; however, we need to know the restrictions of $\theta$ and $\eta$. Thus, using again the expression of the product $\tau\lambda$, we have Table \ref{tab5}, where there are four possible constraints for $\theta$ and $\eta$ depending of the values of $s'$.
\begin{center}
\begin{table}[h]
\centering
\caption{Restrictions for the NC parameters $\theta$ and $\eta$.}
\def\arraystretch{1.1}
\begin{tabular}{cccc}
\hline
Restrictions & $s'$ & $\theta$ & $\eta$ \\ \hline
1 & $+1$ & $\theta>4/m_0 \omega_c$ & $\eta>m_0 \omega_c$ \\
2 & $+1$ & $\theta<4/m_0 \omega_c$ & $\eta<m_0 \omega_c$ \\
3 & $-1$ & $\theta>4/m_0 \omega_c$ & $\eta<m_0 \omega_c$ \\
4 & $-1$ & $\theta<4/m_0 \omega_c$ & $\eta>m_0 \omega_c$ \\ \hline
\end{tabular}
\label{tab5}
\end{table}
\end{center}

\begin{figure}[ht]
\centering
\includegraphics[scale=1.0]{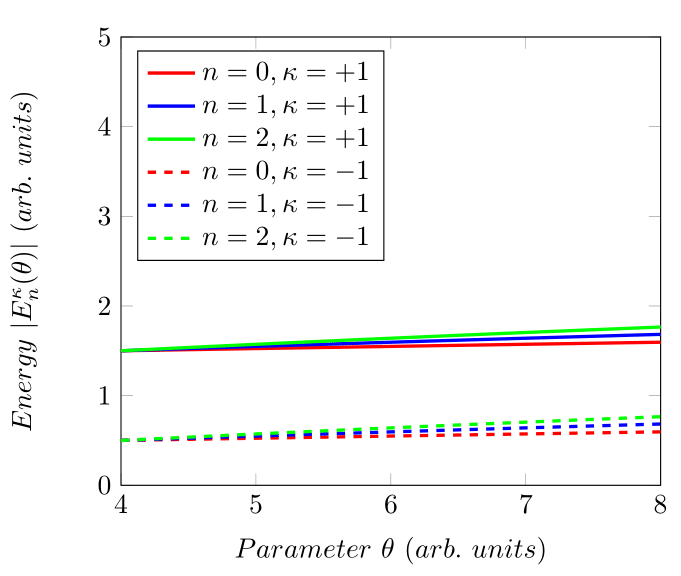}
\caption{Graph of $\vert E^{\kappa}_n(\theta)\vert$ versus $\theta$ for three different values of $n$.}
\label{fig3}
\end{figure}

However, as we use restriction 1 for the field $B$ (see Table \ref{tab3}), it implies that we can only use the restrictions 1 or 2 of Table \ref{tab5}. Besides, choosing a given restriction (here we choose the first for all graphs), then the values of $\theta$ and $\eta$ must be used to construct the graph of $\vert E^{\kappa}_n (\theta)\vert$ versus $\theta$ as well as the graph of $\vert E^{\kappa}_n (\eta)\vert$ versus $\eta$, respectively. In this way, in Fig. \ref{fig3} we see the behavior of the energies of the particle and antiparticle as a function of $\theta$ for three different values of $n$ ($n=0,1,2$), where $m_0=e=a=B=1$, $\eta= 1.1$, and $4<\theta<\infty$. According to this Figure, we see that the function of $\theta$ is to increase (``softly'') the energies of both particle and antiparticle. Furthermore, we see that the energies increase with the increase of $n$ (as it should be) and increase as a function of $B$. So, comparing both the energies of the particle and antiparticle, we see that the energies of the particle are always greater than those of the antiparticle with a difference of one unity, that is: $E^+_n(\theta)=1+\vert E^-_n(\theta)\vert$ (solid lines are parallel with the dashed lines).

\begin{figure}[ht]
\centering
\includegraphics[scale=1.0]{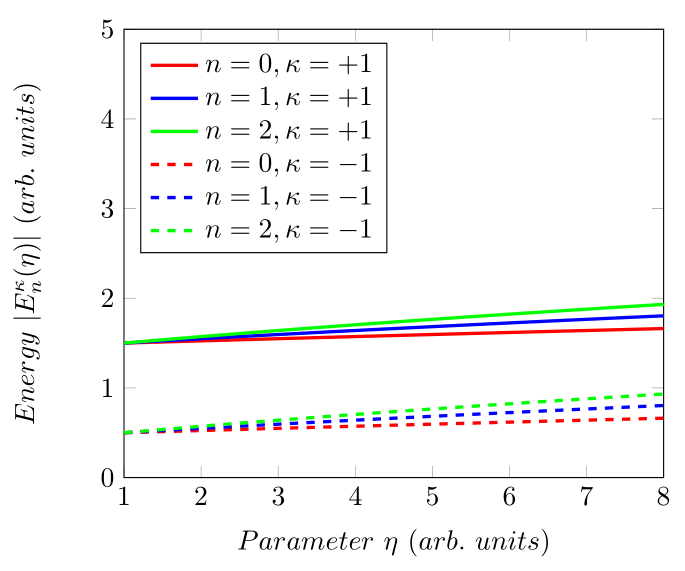}
\caption{Graph of $\vert E^{\kappa}_n(\eta)\vert$ versus $\eta$ for three different values of $n$.}
\label{fig4}
\end{figure}

Already in Fig. \ref{fig4}, we see the behavior of the energies of the particle and antiparticle as a function of $\eta$ for three different values of $n$, where $m_0=e=a=B=1$, $\theta=4.1$, and $1<\theta<\infty$. In particular, the graph this figure is very similar to the graph of $\vert E^{\kappa}_n (\theta)\vert$ versus $\theta$, and therefore, the function of the parameter $\eta$ (and $n$) also is to increase the energies, where the energies of the particle also are always greater than those of the antiparticle with a difference of one unit, that is: $E^+_n(\eta)=1+\vert E^-_n(\eta)\vert$. However, the energy difference between two consecutive energy levels is larger in the graph $E$ versus $\eta$, thus showing that the parameter $\eta$ affects the particle and antiparticle more ``strongly'' than the parameter $\theta$.

Before ending this section, let's now focus on the form of the NC Dirac spinor (inertial flat spinor) for the relativistic bound states of the NCQHE, where such a spinor is symbolized by $\Psi^{NC}_D$, and given by the expression $\Psi^{NC}_D=U\psi^{NC}_C$. So, using the fact that the dimensionless variable $r$ is written as $r=m_0\vert\Omega\vert\rho^2$, implies that we can rewrite the function \eqref{M16} as follows
\begin{equation}\label{M20}
f(\rho)=C(m_0\vert\Omega\vert)^{\frac{\vert\Gamma\vert}{2}}\rho^{\vert\Gamma\vert}e^{-\frac{m_0\vert\Omega\vert\rho^2}{2}}L^{\vert\Gamma\vert}_n(m_0\vert\Omega\vert\rho^2), \ \ \Gamma=\left(m_j-\frac{s}{2}\right).
\end{equation}

To obtain the inferior radial function of the spinor, given by $g(\rho)$, just substitute (\ref{M20}) in Eq. (\ref{M11}), where we get
\begin{equation}\label{M21}
g(\rho)=CD(m_0\vert\Omega\vert)^{\frac{\vert\Gamma\vert}{2}}\rho^{\vert\Gamma\vert}e^{-\frac{m_0\vert\Omega\vert\rho^2}{2}}\left[R(\rho)L^{\vert\Gamma\vert}_{n}(m_0\vert\Omega\vert\rho^2)-2m_0\vert\Omega\vert\rho L^{\vert\Gamma\vert+1}_{n-1}(m_0\vert\Omega\vert\rho^2)\right],
\end{equation}
with
\begin{equation}\label{D1}
D\equiv\frac{\tau}{(m_0+E-E_m)}, \ \ R(\rho)\equiv\left((\vert\Gamma\vert-s\Gamma)\frac{1}{\rho}-(\vert\Omega\vert+s\Omega)m_0\rho\right),
\end{equation}
where $g(\rho)$ must also satisfy the following boundary conditions to be a normalizable solution
\begin{equation}\label{M22} 
g(\rho\to 0)=g(\rho\to\infty)=0.
\end{equation}.

Then, from the radial functions (\ref{M20}) and (\ref{M21}), it implies that the curvilinear spinor (\ref{spinor}) takes the following form
\begin{equation}\label{spinor2}
\psi^{NC}_C=\frac{C(m_0\vert\Omega\vert)^{\frac{\vert\Gamma\vert}{2}}}{\sqrt{2\pi}}e^{i(m_j\varphi-Et)}\rho^{\vert\Gamma\vert}e^{-\frac{m_0\vert\Omega\vert\rho^2}{2}}\left(
           \begin{array}{c}
            L^{\vert\Gamma\vert}_{n}(m_0\vert\Omega\vert\rho^2) \\
            iD\left[R(\rho)L^{\vert\Gamma\vert}_{n}(m_0\vert\Omega\vert\rho^2)-2m_0\vert\Omega\vert\rho L^{\vert\Gamma\vert+1}_{n-1}(m_0\vert\Omega\vert\rho^2)\right] \\
           \end{array}
         \right).
\end{equation}

Therefore, as $\Psi^{NC}_D=e^{-\frac{i\varphi\Sigma_3}{2}}\psi^{NC}_C$, being $\Sigma_3=\sigma_3$ and $e^{-\frac{i\Sigma_3\varphi}{2}}=$diag$(e^{-\frac{i\varphi}{2}},e^{\frac{i\varphi}{2}})$, we get the following NC Dirac spinor
\begin{equation}\label{spinor3} 
\Psi^{NC}_D(t,\rho,\varphi)=\Phi(t,\rho,\varphi)\left(
           \begin{array}{c}
            L^{\vert\Gamma\vert}_{n}(m_0\vert\Omega\vert\rho^2) \\
            iD\left[R(\rho)L^{\vert\Gamma\vert}_{n}(m_0\Omega\rho^2)-2m_0\vert\Omega\vert\rho L^{\vert\Gamma\vert+1}_{n-1}(m_0\vert\Omega\vert\rho^2)\right] \\
           \end{array}
         \right),
\end{equation}
where the function $\Phi(t,\rho,\varphi)$ is defined as follows
\begin{equation}\label{Phi} 
\Phi(t,\rho,\varphi)\equiv\frac{C(m_0\vert\Omega\vert)^{\frac{\vert\Gamma\vert}{2}}}{\sqrt{2\pi}}e^{i(m_l\varphi-Et)}\rho^{\vert\Gamma\vert}e^{-\frac{m_0\vert\Omega\vert\rho^2}{2}},
\end{equation}
and we use the fundamental relationship between the quantum numbers $m_j$, $m_l$ and $m_s$, i.e.: $m_j=m_l+\frac{s}{2}$. In fact, we can do this because $\Psi^{NC}_D(t,\rho,\varphi)$ satisfies the periodicity condition given by: $\Psi^{NC}_D(t,\rho,\varphi\pm 2\pi)=\Psi^{NC}_D(t,\rho,\varphi)$.

Here, it is important to highlight that our Dirac spinor simultaneously incorporates the positive and negative values of the quantum number $m_l$ (or $m_j$), which does not occur, for example, in Refs. \cite{Villalba,Schluter,Villalba2003}. In particular, this happens because we leave implicit (``hidden'') the factor $\sqrt{\rho}$ in the expression of the curvilinear spinor, given by: $\psi^{NC}_C(t,\rho,\varphi)=U^{-1}(\varphi)\Psi^{NC}_D(t,\rho,\varphi)$. In other words, we prefer to let such a factor appear naturally in the final expression of the spinor, where such a factor can be extracted from the term $\rho^{\vert\Gamma\vert}$ as: $\rho^{\vert\Gamma\vert}=\frac{\rho^{\vert m_j\vert}}{\sqrt{\rho}}$. So, although we did not explicitly load the factor $\sqrt{\rho}$ from the beginning, it appears in the final expression of the spinor(s), which does not detract from our problem at all. Furthermore, according to Ref. \cite{Villalba}, such a factor was introduced (conveniently) in the curvilinear spinor expression to cancel a term of the expression $\frac{U\partial_\varphi U^{-1}}{\rho}$, that is given by $\frac{1}{2\rho}$ in (\ref{M9}). However, this could also be done simply by renaming the radial functions as: $f(\rho)\to\frac{1}{\sqrt{\rho}}F(\rho)$ and $g(\rho )\to\frac{1}{\sqrt{\rho}}G(\rho)$ \cite{Villalba2003}. From the above, we followed a simpler path to be able to write the positive and negative values of $m_l$ in a single spinor. From a practical point of view, one of the advantages of having a spinor in the form (\ref{spinor3}) is the possibility of computing the physical observables (expected values) in a faster and more direct way than if we had two spinors, one for each value of $m_l$.

\subsection{Nonrelativistic limit}\label{subsec2}

So, to analyze the nonrelativistic limit (or low-energy limit) of our results, it is necessary to consider that most of the total energy of the system is concentrated in the rest energy of the particle \cite{Greiner}, and whose standard prescription for this is given by: $E\simeq m_0+\varepsilon$, where $m_0\gg \varepsilon$ and $m_0\gg E_m$. Therefore, using this prescription in Eq. (\ref{M13}) with $s=+1$ (spinless), we get the following Schrödinger equation (SE) for the NCQHE with AMM in the usual Euclidean space
\begin{equation}\label{lim1}
i\partial_t\Psi^{NC}_S(t,\rho,\varphi)=H_S^{NC} \Psi^{NC}_S(t,\rho,\varphi)=[H_{NCQHO-like}+H_{Zeeman}^{NC}]\Psi^{NC}_S(t,\rho,\varphi),
\end{equation}
where
\begin{equation}\label{H1}
H_{NCQHO-like}=-\frac{\tau^2}{2m_0}\nabla^2+\frac{1}{2}m_0(\tau\Omega\rho)^{2}=-\frac{\tau^2}{2m_0}\left(\frac{\partial^{2}}{\partial\rho^{2}}+\frac{1}{\rho}\frac{\partial}{\partial\rho}-\frac{L^2_z}{\rho^{2}}\right)+\frac{1}{2}m_0(\tau\Omega\rho)^{2},
\end{equation}
and
\begin{equation}\label{H2}
H_{Zeeman}^{NC}=-\vec{\mu}_{eff}\cdot\vec{B}_{ext},
\end{equation}
with
\begin{equation}\label{lim2}
\vec{\mu}_{eff}=-(a\vec{n}+\tau\lambda\vec{L})\mu_B=-(\vec{\mu}_m-\tau\lambda\vec{\mu}_L),\ \  \vec{\mu}_L=-\mu_B\vec{L},\ \ \vec{L}=(0,0,L_z)=(0,0,-i\partial_\varphi),
\end{equation}
being $H_S^{NC}$ the NC Schrödinger Hamiltonian, where $H_{NCQHO-like}$ is the noncommutative quantum harmonic oscillator (NCQHO)-like Hamiltonian, $H_{Zeeman}^{NC}$ is the NC Zeeman hamiltonian, whose eigenvalue $E_{Zeeman}^{NC}$ is called NC Zeeman energy (another kind of magnetic potential energy), $\Psi^{NC}_S(t,\rho,\varphi)=\frac{e^{i(m_l\varphi-\varepsilon t)}}{\sqrt{2\pi}}\hat{f}(\rho)$, or $\Psi^{NC}_S(t,\rho,\varphi)=\frac{e^{i(m_l\varphi-\varepsilon t)}}{\sqrt{2\pi\rho}}\hat{F}(\rho)$ \cite{Villalba2003}, is the NC Schrödinger wave function, or simply the NC wave function, $\nabla^2$ is the Laplacian operator, $\vec{L}$ is the orbital angular momentum (operator), $\vec{\mu}_{eff}$ is the effective or total MDM of the particle (``classical electron''), where $\vec{\mu}_L=\mu_L\vec{e}_z$ ($\mu_L=\mu_z=-g\mu_{B}m_l; g=1$) is the orbital MDM (MDM associated with the ``orbital'' or cyclotronic motion), $\vec{n}=(0,0,1)$ is a unit vector (entered for convenience) and $\mu_B$ is the Bohr magneton. As we can see in (\ref{lim1}), Schrödinger Hamiltonian has two kinds of MDMs: orbital and anomalous (both of a different nature), or two kinds of magnetic potential energy: one continuous and one quantized, that is, the first and second term in $H^{NC}_{Zeeman}$.

By way of information, the Zeeman effect happens when an atom is placed in a uniform external magnetic field, where the energy levels of the electron(s) are shifted. In other words, it is the effect of splitting a spectral line into several components in the presence of a static magnetic field. Furthermore, in the absence of the NC phase space ($\theta=\eta=0$) and of the AMM ($a=0$), we get exactly the usual Zeeman Hamiltonian, given by: $H_{Zeeman}=-\vec{\mu}_L\cdot\vec{B}_{ext}=\frac{e}{2m_0}\vec{L}\cdot\vec{B}_{ext}$, where $\vec{L}\cdot\vec{B}_{ext}>0$ ($L_z$ and $B_z$ are parallel) is the condition for the maximum energy, and $\vec{L}\cdot\vec{B}_{ext}<0$ ($L_z$ and $B_z$ are antiparallel) is the condition for the minimum energy. Furthermore, in our problem we verified that the function of the AMM as well as the NCs parameters is to increase (to $m_l>0$ and $\tau\lambda>0$, or $m_l<0$ and $\tau\lambda<0$) or decrease (to $m_l>0$ and $\tau\lambda<0$, or $m_l<0$ and $\tau\lambda>0$) the energies of the particle, whose energy spectrum is given by: $E_{Zeeman}^{NC}=E_m+E_{m_l}=\mu_m B+\frac{s'}{2}\vert\tau\lambda\vert\omega_c m_l$. However, we see that if the restrictions for $\tau\lambda$, which are $\tau\neq 0$ and $\lambda\neq 0$, are not obeyed, then we would only have the AMM in the Zeeman Hamiltonian, with $H_S^{NC}=H_{Zeeman}=a\mu_{B}B$, that is, we would have something with the ``anomalous Zeeman Hamiltonian'' (not to be confused with the true anomalous Zeeman Hamiltonian).

So, again using the prescription $ E \simeq m_0 + \varepsilon $ (with $ m_0 \gg \varepsilon $, $ m_0 \gg E_m $ and $s=+1$), now in (\ref{spectrum}), we obtain the following nonrelativistic energy spectrum (nonrelativistic Landau levels) for the NCQHE with AMM in the usual Euclidean space
\begin{equation}\label{spectrum2}
\varepsilon_{n,m_l,s'}=E_{NCQHO-like}+E_{Zeeman}^{NC}=E_m+\vert\tau\lambda\vert\omega_c\left[n+\frac{1}{2}+\frac{\vert m_l\vert+s' m_l}{2}\right],
\end{equation}
where $E_{NCQHO-like}$ are the eigenvalues of $H_{NCQHO-like}$, and we use the following substitution: $\vert m_j-1/2\vert+s'(m_j+1/2)\to \vert m_l\vert+s' m_l$, something consistent from a nonrelativistic point of view, since the particle has no spin and $m_j$ should not appear in the spectrum. In particular, we note that the spectrum (\ref{spectrum2}) has some similarities and some differences with the relativistic case (for the particle). For example, unlike the relativistic case, the spectrum (\ref{spectrum2}) only admits positive energy states ($\varepsilon_{n,m_l,s'}>0$), linearly depends on the quantum numbers $n$ and $m_l$, cyclotron frequency $\omega_c$, and of the NC parameters $\theta$ and $\eta$. Now, similar to the relativistic case, the spectrum (\ref{spectrum2}) linearly depends on the magnetic energy $E_m$, has a finite or infinite degeneracy ($s' m_l>0$ ou $s' m_l\leq 0$), increases as a function of quantum numbers $n$ e $m_l$ (for $s' m_l>0$), and still remains quantized even in the absence of the magnetic field ($B=0$), whose spectrum is given by: $\varepsilon_{n,m_l}=(\eta/m_0)\left[n+\frac{1}{2}+\frac{\vert m_l\vert+m_l}{2}\right]>0$. Furthermore, comparing the spectrum (\ref{spectrum2}) with the literature, we verified that in the absence of the AMM ($E_m=0$) with $s'm_l<0$, we obtain the spectrum of the nonrelativistic QHE in a NC space and in a NC phase space (in the absence of an electric field) \cite{Dulat,Dayi}. Already in the absence of the NC phase space ($\theta=\eta=0$) and of the AMM ($E_m=0$), we get the usual spectrum of the nonrelativistic QHE for all possible values of $m_l$\cite{Dulat,Dayi,Villalba,Yoshioka,Das,Brand,Lima}. From the above, we clearly see that our nonrelativistic spectrum generalizes many particular cases of the literature.

Before ending this section, let's now obtain the NC wave function for the nonrelativistic bound states of the QHE. In particular, this function can be obtained in two different ways (but equivalent), namely: directly solving Eq. (\ref{lim1}) or starting directly from the function (\ref{M20}). For simplicity, we choose this second option. Therefore, using the following substitutions
\begin{equation}\label{substitutions}
\vert\Gamma\vert\to\vert m_l\vert, \ \ C\to C',
\end{equation}
we have
\begin{equation}\label{lim3}
f(\rho)=C'(m_0\vert\Omega\vert)^{\frac{\vert m_l\vert}{2}}\rho^{\vert m_l\vert}e^{-\frac{m_0\vert\Omega\vert\rho^2}{2}}L^{\vert m_l\vert}_n(m_0\Omega\rho^2),
\end{equation}
where $C'>0$ is a new (nonrelativistic) normalization constant.

In this way, the NC wave function takes on the following form
\begin{equation}\label{lim4}
\Psi^{NC}_S(t,\rho,\varphi)=\frac{C'(m_0\vert\Omega\vert)^{\frac{\vert m_l\vert}{2}}}{\sqrt{2\pi}}e^{i(m_l\varphi-\varepsilon t)}\rho^{\vert m_l\vert}e^{-\frac{m_0\vert\Omega\vert\rho^2}{2}}L^{\vert m_l\vert}_n(m_0\vert\Omega\vert\rho^2),
\end{equation}
in which it must satisfy the following boundary conditions to be a normalizable solution
\begin{equation}\label{lim5} 
\Psi^{NC}_S(t,\rho\to 0,\varphi)=\Psi^{NC}_S(t,\rho\to\infty,\varphi)=0,
\end{equation}
and also the following periodicity condition (analogous to the relativistic case)
\begin{equation}\label{periodicidade} 
\Psi^{NC}_S(t,\rho,\varphi\pm 2\pi)=\Psi^{NC}_S(t,\rho,\varphi).
\end{equation}

\section{The noncommutative Dirac equation in a (2+1)-dimensional generic curved spacetime\label{sec4}}

Here, we introduce the NCDE in a (2+1)-dimensional generic curved spacetime. To achieve this goal, we use the tetrad formalism since it is a very efficient method to introduce fermions (or fermionic fields) into curved spacetimes (gravitational fields). So, in polar coordinates $(t,\rho,\varphi)$, the line element for a generic curved spacetime can be written by the following expression
\begin{equation}\label{lineelement1}
ds^2_{generic}=g_{\mu\nu}(x)dx^\mu dx^\nu=\left(X dt+Y d\varphi\right)^{2}-d\rho^{2}-Z^{2}d\varphi^{2}, \ \ (\mu,\nu=t,\rho,\varphi),
\end{equation}
where the coefficients $X$, $Y$ and $Z$ are functions only of the polar radial coordinate: $X=X(\rho)$, $Y=Y(\rho)$ and $Z=Z(\rho)$, and $g_{\mu\nu}(x)$ is the curved metric tensor (or simply curved metric), whose inverse is given $g^{\mu\nu}(x)$, and both take the following form
\begin{equation}\label{metric}
g_{\mu\nu}(x)=\left(\begin{array}{ccc}
X^2 & \ 0 & XY \\
0 & -1 &  0 \\
XY & \ 0 & Y^2-Z^2
\end{array}\right), \ \ 
g^{\mu\nu}(x)=\left(\begin{array}{ccc}
\frac{Z^{2}-Y^{2}}{X^{2}Z^{2}} & 0 & \frac{Y}{XZ^{2}}\\
0 & -1 & 0\\
\frac{Y}{XZ^{2}} & 0 & -\frac{1}{Z^{2}}
\end{array}\right).
\end{equation}

With the line element given by the expression (\ref{lineelement1}), we now need to build a local reference frame where the observer will be placed (laboratory frame). Consequently, it is in this local reference frame that we can then define the gamma matrices (or the spinor) in a curved spacetime \cite{BakkePRD,Oliveira2019,OliveiraGRG,Lawrie}. However, through the tetrad formalism, it is perfectly possible to achieve this objective. In particular, the tetrad formalism states that a given curved spacetime can be introduced point to point with a flat spacetime through objects of the type $e_{\ \ a}^\mu(x)$, which are called tetrads (square matrices), and which together with their inverses, given by $e^a_{\ \ \mu}(x)$, satisfy the following relationships: $\hat{\theta}^a=e^a_{\ \ \mu}(x)dx^\mu$ e $dx^\mu=e_{\ \ a}^\mu(x)\hat{\theta}^a$, where $\hat{\theta}^a$ is a quantity called the a noncoordinate basis. By convention, the index given by the Greek letters refers to curved spacetime, while the Latin letters to Minkowski flat spacetime, respectively. Furthermore, tetrads and their inverses must also satisfy the following relations \cite{BakkePRD,Oliveira2019,OliveiraGRG,Lawrie}. 
\begin{equation}\label{tetrads}
e^{a}_{\ \ \mu}(x)e^\mu_{\ \ b}(x)=\delta^a_{\ \ b}, \ \ e^{\mu}_{\ \ a}(x)e^{a}_{\ \ \nu}(x)=\delta^\mu_{\ \ \nu}, \ \ (\mu,\nu=t,\rho,\varphi; \ a,b=0,1,2),
\end{equation}
and
\begin{equation}\label{tensor}
g_{\mu\nu}(x)=e^a_{\ \ \mu}(x)e^b_{\ \ \nu}(x)\tilde{\eta}_{ab}, \ \ \tilde{\eta}_{ab}=e^\mu_{\ \ a}(x)e^\nu_{\ \ b}(x)g_{\mu\nu}(x),
\end{equation}
where $\tilde{\eta}_{ab}=$diag$(1,-1,-1)$ is the  (non-polar) Cartesian Minkowski metric and $\delta^{a(\mu)}_{\ \ b (\nu)}=$diag$(1,1,1)$ is the (2+1)-dimensional Kronecker delta. Then, through the tetrad formalism, we can rewrite the line element (\ref{lineelement1}) in terms of the noncoordinate basis as follows
\begin{equation}\label{lineelement2}
ds^2_{generic}=g_{\mu\nu}(x)dx^\mu dx^\nu=\tilde{\eta}_{ab}\hat{\theta}^a\otimes\hat{\theta}^b=(\hat{\theta}^0)^2-(\hat{\theta}^1)^2-(\hat{\theta}^2)^2, \ \ (a,b=1,2,3),
\end{equation}
where the components of $\hat{\theta}^{a(b)}$ are written as
\begin{equation}\label{bases}
\hat{\theta}^0=Xdt-Yd\varphi, \ \ \hat{\theta}^1=d\rho, \ \ \hat{\theta}^2=Zd\varphi,
\end{equation}
being $\otimes$ the symbol for the tensor product. As a result, the tetrads and their inverses take the form
\begin{equation}\label{tetrads}
e^{\mu}_{\ \ a}(x)=\left(
\begin{array}{ccc}
 \frac{1}{X} & 0 & -\frac{Y}{XZ} \\
 0 & 1 & 0 \\
 0 & 0 & \frac{1}{Z} \\
\end{array}
\right), \ \
e^{a}_{\ \ \mu}(x)=\left(\begin{array}{ccc}
X & 0 & Y \\
0 & 1 & 0 \\
0 & 0 & Z
\end{array}\right).
\end{equation}

As we can clearly see in (\ref{bases}), where the components of $\hat{\theta}^{a(b)}$ are explicit functions of the polar coordinate, here, a given quantity (or parameter) with the indices $a, b,c,\ldots$ does not necessarily mean that such quantity (or parameter) has Cartesian coordinates (here the metric $\tilde{\eta}_{ab}$ is an ``exception''), as it happens in the inertial flat case (actually Latin indices are used to represent Minkowski spacetime, which can have either Cartesian or polar coordinates). For example, taking $X=1$, $Y=0$ and $Z=\rho$ in (\ref{tensor}), we obtain exactly the polar Minkowski metric, something it would not be able to do if $\tilde{\eta}_{ab}$ were also written in polar coordinates.

Now, we can get one of the fundamental objects of DE in curved spacetimes (in the absence of torsion), where such an object is called a spin connection (antisymmetric tensor) \cite{Chernodub1,Chernodub2,Lawrie}, in which it is defined as follows
\begin{equation}\label{spinconnection}
\omega_{ab\mu}(x)=-\omega_{ba\mu}(x)=\tilde{\eta}_{ac}\omega^c_{\ \ b\mu}(x)=\tilde{\eta}_{ac}e^c_{\ \ \nu}(x)\left[e^\sigma_{\ \ b}(x)\Gamma^\nu_{\ \mu\sigma}(x)+\partial_\mu e^\nu_{\ \ b}(x)\right], 
\end{equation}
where
\begin{equation}\label{Christoffel}
\Gamma^\nu_{\ \mu\sigma}(x)=\Gamma^\nu_{\ \sigma\mu}(x)=\frac{1}{2}g^{\nu\lambda}(x)\left[\partial_\mu g_{\lambda\sigma}(x)+\partial_\sigma g_{\lambda\mu}(x)-\partial_\lambda g_{\mu\sigma}(x)\right], 
\end{equation}
are the Christoffel symbols of the second type (symmetric tensor) and $\omega^c_{\ \ b\mu}(x)$ are sometimes called spin connection coefficients. Explicitly, the non-zero components of the Christoffel symbols are given by
\begin{equation}\label{symbol1}
\Gamma^{t}_{\ t\rho}(x)=\Gamma^{t}_{\ \rho t}(x)=\frac{2Z^2 X'+XYY'-Y^2 X'}{2XZ^2},
\end{equation}
\begin{equation}\label{symbol2}
\Gamma^{t}_{\ \rho\varphi}(x)=\Gamma^{t}_{\ \varphi\rho}(x)=\frac{XY^2 Y'-Y^3 X'+XZ^2 Y'+YZ(ZX'-2XZ')}{2X^2 Z^2},
\end{equation}
\begin{equation}\label{symbol3}
\Gamma^{\rho}_{\ t t}(x)=XX',
\end{equation}
\begin{equation}\label{symbol4}
\Gamma^{\rho}_{\ t \varphi}(x)=\Gamma^{\rho}_{\ \varphi t}(x)=\frac{1}{2}(YX'+XY'),
\end{equation}
\begin{equation}\label{symbol5}
\Gamma^{\rho}_{\ \varphi\varphi}(x)=YY'-ZZ',
\end{equation}
\begin{equation}\label{symbol6}
\Gamma^{\varphi}_{\ t\rho}(x)=\Gamma^{\varphi}_{\ \rho t}(x)=\frac{YX'-XY'}{2Z^2},
\end{equation}
\begin{equation}\label{symbol7}
\Gamma^{\varphi}_{\ \rho\varphi}(x)=\Gamma^{\varphi}_{\ \varphi\rho}(x)=\frac{Y^2 X'-XYY'+2XZZ'}{2XZ^2},
\end{equation}
where $X'=\frac{dX}{d\rho}$, $Y'=\frac{dY}{d\rho}$, and $Z'=\frac{dZ}{d\rho}$, respectively.

Consequently, the non-zero components of the spin connection are written as
\begin{equation}\label{component1}
\omega_{01t}(x)=-\omega_{10t}(x)=-X',
\end{equation}
\begin{equation}\label{component2}
\omega_{12t}(x)=-\omega_{21t}(x)=\frac{XY'-YX'}{2C},
\end{equation}
\begin{equation}\label{component3}
\omega_{02\rho}(x)=-\omega_{20\rho}(x)=\frac{XY'-YX'}{2XZ},
\end{equation}
\begin{equation}\label{component4}
\omega_{01\varphi}(x)=-\omega_{10\varphi}(x)=-\frac{XY'+YX'}{2X},
\end{equation}
\begin{equation}\label{component5}
\omega_{12\varphi}(x)=-\omega_{21\varphi}(x)=\frac{Y(XY'-YX')}{2XZ}-Z'.
\end{equation}

From now on, we will focus our attention on the NCED in a generic curved spacetime. Thus, we have the following tensorial DE with minimal and nonminimal couplings in a generic curved spacetime (in curvilinear coordinates) \cite{Oliveira2019,GRG,OliveiraGRG,Matsuo1}
\begin{equation}\label{dirac1}
\left\{\gamma^\mu(x)[P_\mu(x)-qA_\mu (x)]+\frac{\mu_{m}}{2}\sigma^{\mu\nu}(x)F_{\mu\nu}(x)-m_0\right\}\psi_C(t,\rho,\varphi)=0, \ \ (\mu,\nu=t,\rho,\varphi),
\end{equation}
where $\gamma^{\mu}(x)=e^\mu_{\ \ a}(x)\gamma^a$ are the curved gamma matrices, which satisfy the anticommutation relation of the covariant Clifford algebra: $\{\gamma^\mu(x),\gamma^\nu(x)\}=2g^{\mu\nu}(x)I_{2\times 2}$, $P_\mu(x)=i\nabla_\mu (x)=i[\partial_\mu+\Gamma_\mu (x)]$ is the curved moment operator, $\nabla_\mu(x)$ is the covariant derivative, with $\partial_\mu$ being the usual partial derivatives ($\partial_\mu\neq e^c_{\ \mu} (x)\partial_c$), $\Gamma_\mu(x)=-\frac{i}{4}\omega_{cb\mu}(x)\sigma^{cb}$ is the spinorial connection (or spinor affine connection), $\sigma^{\mu\nu}(x)=\frac{i}{2}[\gamma^\mu(x),\gamma^\nu(x)]$ is a curved antisymmetric tensor, $F_{\mu\nu}(x)=F_{\mu\nu}(x)=\partial_\mu A_\nu (x)-\partial_\nu A_\mu (x)=e^a_{\ \ \mu}(x)e^b_{\ \ \nu}(x)F_{ab}$ is the curved electromagnetic field tensor, with $A_\mu (x)=e^b_{\ \ \mu} (x)A_{b}$ being the curved electromagnetic potential (field), and $\psi_C(t,\rho,\varphi)=e^{\frac{i\varphi\Sigma^3}{2}}\Psi_D(t,\rho,\varphi)$ is our curvilinear spinor, where $\Psi_D(t,\rho,\varphi)$ is our original Dirac spinor \cite{Schluter,Villalba,Villalba2003}. As we saw in the inertial flat case, this exponential factor appears in the spinor when we convert the Cartesian DE into a curvilinear DE.

In particular, the product $\gamma^{\mu}(x) A_\mu(x)$ is equal to the result of the inertial flat case with a ``correction'' in the component  $A_2$ of the electromagnetic potential, that is
\begin{equation}\label{potential}
\gamma^\mu(x)A_\mu(x)=e^\mu_{\ \ a}(x) e^b_{\ \ \mu}(x)\gamma^a A_b=\delta_{\ \ a}^b\gamma^a A_b=\gamma^a A_a,
\end{equation}
where $A_a=(A_0,A_1,A_2)=(0,0,-A_\varphi)$. Here, the component $A_2$ must be equal to the angular component $-A_\varphi$ and not $-A_y$ (the negative sign is because of the signature of the metric) \cite{Greiner}, otherwise the inertial curved case would not reduce to the inertial flat case. Furthermore, as we commented above, here a quantity (or parameter) with the indices $a,b,c,\ldots$ does not necessarily mean that such quantity (or parameter) has Cartesian coordinates. On the other hand, analogous to the minimal coupling, the product $\sigma^{\mu\nu}(x)F_{\mu\nu}(x)$ is also equal to the result of the inertial flat case (but with no term of ``correction''), that is
\begin{eqnarray}\label{nonminimal}
\sigma^{\mu\nu}(x)F_{\mu\nu}(x) &=& i\gamma^\mu(x)\gamma^\nu(x)F_{\mu\nu}(x)\nonumber \\
&=& i(e^\mu_{\ \ a}(x) e^c_{\ \ \mu}(x))(e^\nu_{\ \ b}(x)e^d_{\ \ \nu}(x))\gamma^a \gamma^bF_{cd}\nonumber \\
&=& i(\delta^c_{\ \ a})(\delta^d_{\ \ b})\gamma^a \gamma^bF_{cd}\nonumber \\
&=& \sigma^{ab}F_{ab},
\end{eqnarray}
where we get
\begin{equation}
\frac{\mu_m}{2}\sigma^{\mu\nu}(x)F_{\mu\nu}(x)=-2\mu_m\vec{S}\cdot\vec{B}, \ \ \left(\vec{S}=\frac{1}{2}\vec{\Sigma}\right).
\end{equation}

Now, let's introduce the NC phase space in Eq. (\ref{dirac1}). So, starting from the fact that $A_\mu(x)=(0,A_{i'}(x))$, where $A_{i'}(x)=-\frac{B}{2}\epsilon_ {i'j'}x^{j'}(x)$ ($i',j'=\rho,\varphi$) \cite{Szabo}, implies that we can rewrite Eq. (\ref{dirac1}) as follows
\begin{equation}\label{dirac2}
\left\{\gamma^t(x)P_t(x)+\gamma^{i'}(x)\left[P_{i'}(x)+\frac{qB}{2}\epsilon_{i'j'}x^{j'}(x)\right]-2\mu_m\vec{S}\cdot\vec{B}-m_0\right\}\psi_C=0,
\end{equation}
or in an NC phase space, as
\begin{equation}\label{dirac3}
\left\{\gamma^t(x)P_t(x)+\gamma^{i'}(x)\left[P^{NC}_{i'}(x)+\frac{qB}{2}\epsilon_{i'j'}(x^{j'}(x))^{NC}\right]-2\mu_m\vec{S}\cdot\vec{B}-m_0\right\}\star\psi_C=0.
\end{equation}
where we have done $P_{i'}(x)\to P^{NC}_{i'}(x)$, $x^{j'}(x)\to (x^{j'}(x))^{NC}$ and $\psi_C\to\star\psi_C$.

However, to obtain the explicit form of the NCDE in a generic curved spacetime, given by (\ref{dirac3}), we also need to write the NC operators $x^{NC}_\mu$ and $p^{NC} _\mu$ (see (\ref{NC9}) please) in the generic curved spacetime (where $p_\mu\to P_\mu(x)$). For convenient choice, these operators can be written as follows
\begin{equation}\label{dirac4}
(x^\mu(x))^{NC}=x^\mu(x)-\frac{1}{2}\theta^{\mu\nu}P_\nu(x), \ \ P^{NC}_\mu (x)=P_\mu(x)+\frac{1}{2}\eta_{\mu\nu}x^\nu(x), \ \ (\mu,\nu=t,i',j',l'),
\end{equation}
or (NC only in the spatial components)
\begin{equation}\label{dirac5}
(x^{j'}(x))^{NC}=x^{j'}(x)-\frac{1}{2}\theta\epsilon^{j'l'}P_{l'}(x), \ \ P^{NC}_{i'} (x)=P_{i'}(x)+\frac{1}{2}\eta\epsilon_{i'j'}x^{j'}(x).
\end{equation}

Consequently, Eq. (\ref{dirac3}) becomes
\begin{equation}\label{dirac6}
\left\{\gamma^t(x)P_t(x)+\gamma^{i'}(x)\left[\tau P_{i'}(x)-q\lambda A_{i'}(x)\right]-2\mu_m\vec{S}\cdot\vec{B}-m_0\right\}\psi^{NC}_C=0,
\end{equation}
where it results
\begin{eqnarray}\label{dirac7}
    \left[i\gamma^t(x)\partial_t+i\tau\gamma^\rho(x)\partial_\rho+i\tau\gamma^\varphi(x)\partial_\varphi-\frac{\lambda m_0\omega_c}{2}\rho\gamma^2-\mu_m\Sigma^3 B-m_0
\right]\psi^{NC}_C 
 \nonumber \\
    \label{eq:temporal-difference}
    +i[\gamma^t(x)\Gamma_t(x)+\tau\gamma^{\rho}(x)\Gamma_{\rho}(x)+\tau\gamma^{\varphi}(x)\Gamma_{\varphi}(x)]\psi^{NC}_C=0,
\end{eqnarray}
where $\tau=(1-m_0\omega_c\theta/4)$ and $\lambda=(1-\eta/m_0\omega_c)$, being $\omega_c=eB/m_0>0$ ($q=-e<0$), and we use the relation $\gamma^{i'}(x)A_{i'}(x)=\gamma^i A_i=\gamma^2 A_2=-\gamma^2 A_\varphi=-\frac{1}{2}B\rho\gamma^2$.

Furthermore, using the expressions (\ref{component1})-(\ref{component5}), we obtain the following components of the spinorial connection
\begin{equation}\label{spinorial1}
\Gamma_t(x)=\frac{(YX'-XY') \gamma^1\gamma^2+2ZX'\gamma^0\gamma^1}{4Z},
\end{equation}
\begin{equation}\label{spinorial2}
\Gamma_\rho(x)=\frac{(YX'-XY')}{4XZ}\gamma^0\gamma^2, 
\end{equation}
\begin{equation}\label{spinorial3}
\Gamma_\varphi(x)=\frac{(Y^2 X'-XYY'+2XZZ')\gamma^1\gamma^2+(X'YZ+XY'Z)\gamma^0\gamma^1}{4XZ}.
\end{equation}

With respect to curved gamma matrices, we also have
\begin{equation}\label{gamma0}
\gamma^t(x)=\frac{(Z\gamma^0-Y\gamma^2)}{XZ}, 
\end{equation}
\begin{equation}\label{gamma1}
\gamma^\rho(x)=\gamma^1,
\end{equation}
\begin{equation}\label{gamma2}
\gamma^\varphi(x)=\frac{1}{Z}\gamma^2.
\end{equation}

Then, using the expressions (\ref{spinorial1})-(\ref{spinorial3}) and (\ref{gamma0})-(\ref{gamma2}), we obtain the following contribution of the spinorial connection (or spin) to the NCDE
\begin{eqnarray}\label{spinorial}
    [\gamma^t(x)\Gamma_t(x)+\tau\gamma^{\rho}(x)\Gamma_{\rho}(x)+\tau\gamma^{\varphi}(x)\Gamma_{\varphi}(x)]=\frac{(2\tau-1)XY'+YX'}{4XZ}\gamma^{0}\gamma^1\gamma^2
 \nonumber \\
    \label{eq:temporal-difference}
    +\frac{(1-\tau)(XYY'-Y^2X')+2(\tau XZZ'+X'Z^2)}{4XZ^2}\gamma^{1}.
\end{eqnarray}

Therefore, substituting the curved gamma matrices as well as the contribution of the spinorial connection in Eq. (\ref{dirac7}), we have the following NCDE in a generic curved spacetime as a function of the coefficients $X$, $Y$ and $Z$ (coefficients of the line element, or of the metric)
\begin{eqnarray}\label{dirac8}
    \left[i\frac{(Z\gamma^0-Y\gamma^2)}{XZ}\partial_t+i\tau\gamma^1\partial_\rho+i\tau\frac{1}{Z}\gamma^2\partial_\varphi-\frac{\lambda m_0\omega_c}{2}\rho\gamma^2-\mu_m\Sigma^3 B-m_0
\right]\psi^{NC}_C 
 \nonumber \\
    \label{eq:temporal-difference}
    +\left[\frac{(1-\tau)(XYY'-Y^2X')+2(\tau XZZ'+X'Z^2)}{4XZ^2}i\gamma^{1}+\frac{(2\tau-1)XY'-YX'}{4XZ}i\gamma^{0}\gamma^1\gamma^2\right]\psi^{NC}_C=0.
\end{eqnarray}

On the other hand, using the standard ansatz (\ref{spinor}) we can simplify Eq. \eqref{dirac8}, in which we obtain the following time-independent NCDE
\begin{eqnarray}\label{dirac9}
    \left[\frac{(Z\gamma^0-Y\gamma^2)}{XZ}E+i\tau\gamma^1\partial_\rho-\frac{\tau m_j}{Z}\gamma^2-\frac{\lambda m_0\omega_c}{2}\rho\gamma^2-\mu_m\Sigma^3 B-m_0
\right]\phi^{NC}
 \nonumber \\
    \label{eq:temporal-difference}
    +\left[\frac{(1-\tau)(XYY'-Y^2X')+2(\tau XZZ'+X'Z^2)}{4XZ^2}i\gamma^{1}+\frac{(2\tau-1)XY'-YX'}{4XZ}i\gamma^{0}\gamma^1\gamma^2\right]\phi^{NC}=0,
\end{eqnarray}
where $\phi^{NC}(\rho)=(f(\rho),i g(\rho))^T$.

\section{The noncommutative Dirac equation in the spinning cosmic string spacetime\label{sec5}}

In polar coordinates $(t,\rho,\varphi)$, the line element for the spinning CS spacetime can be given by the following expression \cite{Mazur,MSCunha,Cunha,Muniz}
\begin{equation}\label{CCgirante}
ds_{CS}^2=(dt+\beta d\varphi)^2-d\rho^2-\alpha^2\rho^2 d\varphi^2,
\end{equation}
where the Riemann (and Ricci) curvature tensor is written as
\begin{equation}\label{curvatura}
R^{\rho\varphi}_{\rho\varphi}=R^{\rho}_{\rho}=R^{\varphi}_{\varphi}=2\pi\left(\frac{1-\alpha}{\alpha}\right)\delta^2(\vec{r}),
\end{equation}
with $\alpha=1-4\bar{M}$ ($0<\alpha\leq 1$) being the topological or curvature parameter (dimensionless), whose curvature is located on the symmetry axis of the CS ($z$-axis), $\beta=4\bar{J}$ ($\beta\geq 0$) is the rotational or rotation parameter (length dimension), and $\delta^2(\vec{r})$ is the two-dimensional Dirac delta in flat spacetime. With respect to the origin of the parameter $\alpha$, such a parameter arises because the presence of a CS introduces an angular deficit in Minkowski spacetime (``cuts spacetime in the shape of a wedge''), given by: $\Delta\varphi=2\pi(1-\alpha)=\frac{8\pi G\bar{M}}{c^2}$, where such a deficit is a solution of Einstein's equations to CSs. Therefore, from a geometric point of view, CS spacetime is sometimes called Minkowski spacetime with a conical curvature (or conical singularity). In addition, the origin of the parameter $\beta$ is due to the energy-momentum tensor (non-null) for the three-dimensional ``Kerr solution'', therefore, we can define $\bar{J}$ as: $\bar{J}\equiv\frac{1}{2}\epsilon_{ij}\bar{J}^{ij}=\frac{1}{2}\epsilon_{ij}\int d^2 x (x^i T^{0j}-x^j T^{0i})\neq 0$ \cite{Mazur}. Also, for $\alpha\to 1$, it automatically implies in $\beta\to 0$, that is, the absence (or far away) of the CS. However, for $\beta\to 0$, it does not imply in $\alpha\to 1$; in this case, we would have the line element of the static CS, modeled only by $\alpha$. It is also worth mentioning that the line element of the spinning CS is particularly interesting because it can support (for $\rho<\beta/\alpha$) the existence of CTCs, and, consequently, violate causality (or allow time travel). So, to get around this ``problem'', in many problems in the literature, it is assumed that the particle ``orbits'' (moves) far enough away from regions with CTCs (to $\rho>\beta/\alpha$) that exist around the CS (we adopt this condition here) \cite{Muniz,Mazur}.

Therefore, comparing (\ref{CCgirante}) with (\ref{lineelement1}), we get the following coefficients
\begin{equation}\label{coefficients1}
X=1, \ \ Y=\beta, \ \ Z=\alpha\rho,
\end{equation}
and substituting (\ref{coefficients1}) in (\ref{dirac9}), we have the following NCDE in the spinning CS spacetime
\begin{equation}\label{C1}
\left[\gamma^0 E+i\tau\gamma^1\left(\partial_\rho+\frac{1}{2\rho}\right)-\left(\frac{\tau m_j+\beta E}{\alpha\rho}+\frac{\lambda m_0\omega_c}{2}\rho\right)\gamma^2-\mu_m\Sigma^3 B-m_0
\right]\phi^{NC}=0.
\end{equation}

Now, we have to obtain the NCDE in ``quadratic form'', that is, a second-order differential equation for an of the radial functions of the spinor. Analogous to the inertial flat case, this is done by ``separating'' the NCDE (\ref{C1}) into a set of two first-order coupled differential equations. Therefore, using the form of the gamma matrices $\gamma^a=\tilde{\eta}^{ab}\gamma_{b}$ and the matrix $\Sigma^3=\Sigma_3$ as well as the spinor $\phi^{NC}$, implies that these two coupled differential equations are given by
\begin{equation}\label{C2}
\frac{(m_0+E_m-E)}{\tau}f(\rho)=\left[\frac{d}{d\rho}+sm_0\Omega\rho+\frac{s}{\rho}\left(\hat{m}_j+\frac{s}{2}\right)\right]g(\rho),
\end{equation}
\begin{equation}\label{C3}
\frac{(m_0-E_m+E)}{\tau}g(\rho)=\left[\frac{d}{d\rho}-sm_0\Omega\rho-\frac{s}{\rho}\left(\hat{m}_j-\frac{s}{2}\right)\right]f(\rho),
\end{equation}
where $\hat{m}_j\equiv\frac{1}{\alpha}(m_j+\frac{\beta E}{\tau})\neq 0$ is the effective total magnetic quantum number, $\Omega=\frac{\lambda\omega_c}{2\tau}$ is the well-known effective angular frequency, and $E_m=\mu_m B$ is the well-known magnetic energy.

Therefore, substituting (\ref{C3}) into (\ref{C2}), we obtain the following second-order differential DE (``quadratic DE'') for the NCQHE with AMM in spinning CS spacetime
\begin{equation}\label{C4}
\left[\frac{d^2}{d\rho^2}+\frac{1}{\rho}\frac{d}{d\rho}-\frac{\hat{\Gamma}^2}{\rho^2}-(m_0\Omega\rho)^2+\hat{\mathcal{E}}\right]f(\rho)=0,
\end{equation}
where we define
\begin{equation}\label{C5}
\hat{\Gamma}\equiv\left(\hat{m}_j-\frac{s}{2}\right), \ \ \hat{\mathcal{E}}\equiv\frac{(E-E_m)^2-m_0^2}{\tau^2}-2m_0\Omega\left(\hat{m}_j+\frac{s}{2}\right).
\end{equation}

In particular, in the absence of the spinning CS ($\alpha=1$ and $\beta=0$) we obtain exactly the NCDE in the Minkowski spacetime, given by (\ref{M13}).

\subsection{Bound-state solutions: two-component Dirac spinor and relativistic Landau levels}\label{subsec1}

So, to analytically solve Eq. (\ref{C4}), we can follow the entire procedure done in section 3.1. However, due to the similarity of this equation with Eq. (\ref{M13}), we can simply replace $\Gamma$ by $\hat{\Gamma}$ and $m_j$ by $\hat{m}_j$ in the results of Eq. (\ref{M13}) and then we can (easily) get all the results for the inertial curved case. Thus, as the quantization condition of the previous case requires that: $\frac{\vert\Gamma\vert+1}{2}-\frac{\mathcal{E}}{4m_0\vert\Omega\vert}=-n$ ($n=n_\rho=0,1,2,\ldots$), implies that our new quantization condition becomes
\begin{equation}\label{C6}
\frac{\vert\hat{\Gamma}\vert+1}{2}-\frac{\hat{\mathcal{E}}}{4m_0\vert\Omega\vert}=-n,
\end{equation}
where we have the following quadratic polynomial equation for the relativistic total energy $E$
\begin{equation}\label{C7}
E^2-(2 E_m)E-\frac{m_0\vert\tau\lambda\vert\omega_c}{\alpha}\left[s'\left(l+\frac{\beta E}{\tau}\right)+\Big\vert l+\frac{\beta E}{\tau}\Big\vert\right]+[E^2_m-m^2_0-m_0\vert\tau\lambda\vert\omega_c(2n+1+ss')]=0,
\end{equation}
being $l=l(\alpha)\equiv m_j-\frac{s\alpha}{2}$ a ``topological quantum number'' (because it depends on the parameter $\alpha$).

Therefore, analyzing the polynomial equation (\ref{C7}) for $l>0$ and $l<0$ with $s'=\pm 1$, we obtain the following relativistic spectrum for the NCQHE with AMM in the spinning CS spacetime
\begin{equation}\label{spectrum3}
E^{\kappa}_{n,l,s,s'}=\left[E_\alpha+E_m\right]+\kappa\sqrt{\left[E_\alpha+E_m\right]^2-E^2_m + m_0^2+2m_0\vert\tau\lambda\vert\omega_c N_{\alpha}},
\end{equation}
where
\begin{equation}\label{N1}
E_\alpha\equiv s'm_0\omega_c\vert\lambda\vert\beta\left(\frac{\vert l \vert+s'l}{2\alpha l}\right)\gtrless 0, \ N_\alpha\equiv\left(n+\frac{1}{2}+\frac{\vert l \vert+s'l+ss'\alpha}{2\alpha}\right)\geq 0,
\end{equation}
where $\kappa=\pm 1$ is the known well energy parameter (particle/antiparticle), $N_\alpha$ is a ``topological effective quantum number'', and $E_\alpha$ is a ``topological energy'', which only exists because of the field, i.e., for $B=0$ we would have $E_\alpha=0$. In particular, in the absence of the spinning CS ($\alpha=1$ and $\beta=0$), we obtain exactly the spectrum in the Minkowski spacetime, given by (\ref{spectrum}). So, analyzing the spectrum (\ref{spectrum3}), we see that it behaves differently depending on the values of $l$ and $s'$. In fact, we have four possible configurations for the spectrum depending on the values of $l$ and $s'$, as shown in Table \ref{tab6}.
\begin{center}
\begin{table}[h]
\centering
\caption{Energy spectra for the particle and antiparticle.}
\def\arraystretch{1.1}
\begin{tabular}{cccc}
\hline
Configuration & Energy spectrum $E^{\kappa}_{n,l,s,s'}$ & $l$ & $s'$ \\ \hline
1 &$E^{\kappa}_{n,l,s,s'}=[\vert E_\alpha\vert+E_m]+\kappa\sqrt{[\vert E_\alpha\vert+E_m]^2-E_m^2+m_0^2+2m_0\vert\tau\lambda\vert\omega_c N_+}$ & $l>0$ & $+1$ \\
2 &$E^{\kappa}_{n,l,s,s'}=E_m+\kappa\sqrt{m_0^2+2m_0\vert\tau\lambda\vert\omega_c \bar{N}_-}$ & $l>0$ & $-1$ \\
3 &$E^{\kappa}_{n,l,s,s'}=E_m+\kappa\sqrt{m_0^2+2m_0\vert\tau\lambda\vert\omega_c \bar{N}_+}$ & $l<0$ & $+1$ \\
4 &$E^{\kappa}_{n,l,s,s'}=[-\vert E_\alpha\vert+E_m]+\kappa\sqrt{[-\vert E_\alpha\vert+E_m]^2-E_m^2+m_0^2+2m_0\vert\tau\lambda\vert\omega_c N_-}$ & $l<0$ & $-1$ 
\\ \hline
\end{tabular}
\label{tab6}
\end{table}
\end{center}

According to Table \ref{tab3}, we see that for $l>0$ with $s'=+1$ or $l<0$ with $s'=-1$, the spectrum depends on both parameters $\alpha$ and $\beta$, where $N_\pm=\left(n+\frac{1}{2}+\frac{2\vert l \vert\pm s\alpha}{2\alpha}\right)$ and $m_j$ must satisfy the following conditions: $m_j>\frac{s\alpha}{2}$ ($l>0$), which is clearly satisfied for $m_j\geq 1/2 $, or $m_j<\frac{s\alpha}{2}$ ($l<0$), which is clearly satisfied for $m_j\leq -1/2$. Also, due to the presence of the parameter $\alpha$ (not of $\beta$), the degeneracy of the spectrum is broken (``not is well defined''), that is, the conical curvature of the CS breaks the degeneracy of the Landau levels \cite{Lima}. In particular, this is due to the fact that we can no longer construct a third quantum number (integer and positive) from $n$ and $m_l$ ($\vert l\vert/\alpha$ is not an integer) \cite{Rubens}. Already for $l>0$ with $s'=-1$ ($m_j>s\alpha/2$) or $l<0$ with $s'=+1$ ($m_j<s\alpha/2$), the spectrum no longer depends on $\alpha$ and $\beta$ (but it is infinitely degenerate), where $\bar{N}_\pm=\left(n+\frac{1\pm s}{2}\right)$. In this case, it is as if the NCQHE ``lives in the Minkowski spacetime''. However, comparing the spectra of Table \ref{tab6} for $s'=+1$, we see that the spectrum with the highest energies is for the case $l>0$ (increases as a function of $n$ and $m_j$). Now, comparing the spectra for $s'=-1$, we see that the spectrum with the highest energies is for the case $l<0$ (increases as a function of $n$ and $m_j$). Furthermore, the energies of the particle and antiparticle for these two cases are larger when both have spin up ($s=+1$). In particular, the spectrum (\ref{spectrum3}), or configs. 1 and 4, still remains quantized even in the absence of the magnetic field (analogous to the inertial flat case), which is given by: $E^{\kappa}_{n,l,s}=\kappa\sqrt{m_0^2+2\eta N_\alpha}$. As we see, this spectrum does not have the contribution of $\beta$, only the contribution of $\alpha$.

\begin{figure}[ht]
\centering
\includegraphics[scale=1.0]{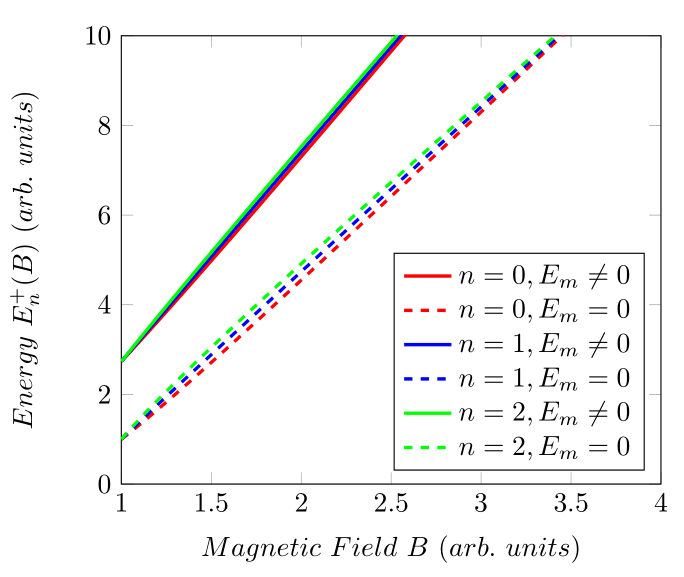}
\caption{Graph of $E^+_n(B)$ versus $B$ for three different values of $n$ with $E_m\neq 0$ ($a=1$) and $E_m=0$ ($a=0$).}
\label{fig5}
\end{figure}

Now, let's analyze the behavior of the spectrum \eqref{spectrum3} as a function of the magnetic field and of the parameters $\alpha$ and $\beta$ for different values of $n$ (with $m_j$ fixes). In this way, we can consider the spectrum of configs. 1 or 4 of Table \ref{tab6}. For simplicity, we chose the config. 1 (also for all graphs here), since we already used $s'=+1$ in the inertial flat case. Therefore, using the restriction 1 of Table \ref{tab3} we have Fig. \ref{fig5}, where it shows the behavior of the energies of the particle as a function of the magnetic field for the ground state ($n=0$) and the first two excited states ($n=1,2$), with and without the presence of magnetic energy, in which we take $m_0=e=a=\theta=\eta=\beta=1$, $m_j=1/2$ and $\alpha=1/2$, being the field given by $1<B<4$. Here, we purposely choose $\alpha=1/2$ for that the spectrum does not depend on spin ($s$ is canceled in the spectrum). In this way, the spectra of the particle (or antiparticle) with spin up or down are exactly the same regardless of the spin chosen.

\begin{figure}[ht]
\centering
\includegraphics[scale=1.0]{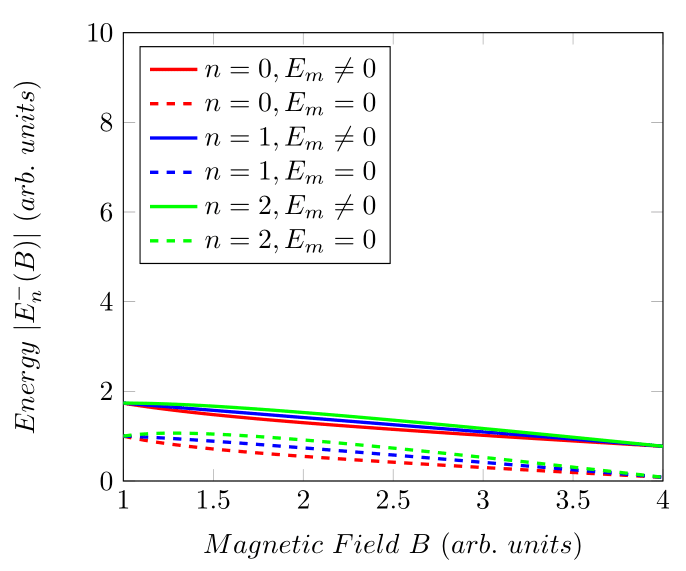}
\caption{Graph of $\vert E^{-}_n (B)\vert$ versus $B$ for three different values of $n$ with $E_m\neq 0$ ($a=1$) and $E_m=0$ ($a=0$).}
\label{fig6}
\end{figure}

According to \ref{fig5}, we see that the energies increase with the increase of $n$ (as it should be), and the function of the AMM as well as of $B$ is to increase the energies of the particle. However, unlike the inertial flat case (see Fig. \ref{fig1}), here, the energies always increase as a function of $B$, and their values are much higher. On the other hand, at $B\approx 1$, the energies with $E_m=0$ are equal for both cases, while the energies with $E_m\neq 0$ are higher for the inertial curved case, respectively. Already in Fig. \ref{fig6}, we have the case of the antiparticle, where we see that the energies increase with the increase of $n$ (as it should be); however, they practically decrease with the increase of $B$ ($E_{final}<E_{initial}$), and for $E_m\neq 0$, are almost null at $B\approx 4$ ($0<\vert E^-_n\vert\ll 1$), where the magnetic energy is almost equal to the quantized energy (``square root energy''). On the other hand, unlike the inertial flat case, here, the energies of the antiparticle are higher in the presence of the AMM ($E_m\neq 0$), i.e., here, the function of the AMM is to increase the energies of the antiparticle (as well as the particle).

\begin{figure}[ht]
\centering
\includegraphics[scale=1.0]{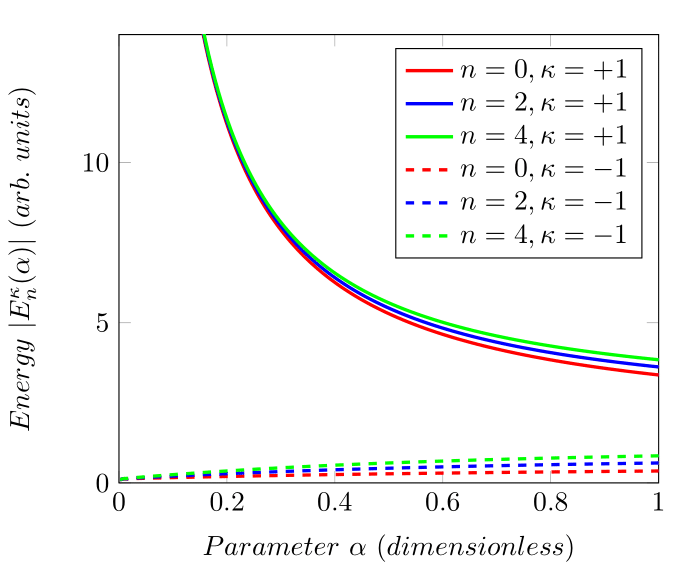}
\caption{Graph of $\vert E^{\kappa}_n(\alpha)\vert$ versus $\alpha$ for three different values of $n$.}
\label{fig7}
\end{figure}

Now, let's analyze the behavior of the spectrum as a function of the parameters $\alpha$ and $\beta$ for different values of $n$ (with $m_j$ fixes). First, let's start by analyzing the graph $\vert E^{\kappa}_n (\alpha)\vert$ versus $\alpha$, and then the graph $\vert E^{\kappa}_n (\beta)\vert$ versus $\beta$, respectively. Therefore, we have Fig. \ref{fig7}, where it shows the behavior of the energies of the particle and antiparticle as a function of $\alpha$ for the ground state ($n=0$), the third excited state ($n=2$) and the fifth excited state ($n=4$), in which we take $m_0=e=a=B=\beta=1$, $\theta=5$, $\eta=2$, and $m_j=1/2$. According to this figure, we see that the energies increase with the increase of $n$, however, are practically equals for $\alpha\leq 0.2$ (particle) and $\alpha\leq 0.1$ (antiparticle). Besides, the energies can increase or decrease while $\alpha$ decreases (an increase of the curvature), i.e., the function of $\alpha$ is to increase the energies of the particle and decrease those of the antiparticle, where the energies of the particle are always much greater than those of the antiparticle. However, unlike the antiparticle, where the energy almost tends to zero at the limit $\alpha\to 0$ (``infinite curvature''), in the case of the particle, the energy tends to ``infinity'' at the limit $\alpha\to 0$, and therefore, the influence of $\alpha$ is much more significant in the case of the particle.

\begin{figure}[ht]
\centering
\includegraphics[scale=1.0]{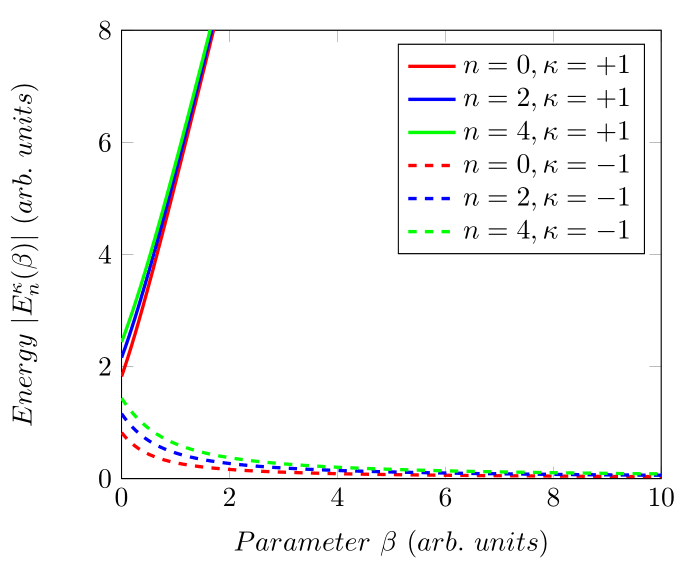}
\caption{Graph of $\vert E^{\kappa}_n(\beta)\vert$ versus $\beta$ for three different values of $n$.}
\label{fig8}
\end{figure}

Already in Fig. \ref{fig8}, we see the behavior of the energies of the particle and antiparticle as a function of $\beta$ for the ground state ($n=0$), the third excited state ($n=2$) and the fifth excited state ($n=4$), where we take $m_0=e=a=B=1$, $\theta=5$, $\eta=2$, $m_j=1/2$ and $\alpha=1/2$. In particular, the function of $n$ and $\beta$ is to increase the energies of the particle (we have $E\to\infty$ for $\beta\to\infty$), and decrease than those of the antiparticle (we have $E\to 0$ for $\beta\to\infty$). In that way, the faster the cosmic string spins, the higher are the energies of the particle and the lower are the energies of the antiparticle.

Before ending this section, let’s now get the NC Dirac spinor (inertial curved spinor) for the relativistic bound states of the NCQHE. So, replacing $\Gamma$ by $\hat{\Gamma}$ in (\ref{spinor2}), we have the following curvilinear spinor for the inertial curved case
\begin{equation}\label{spinor4} 
\psi^{NC}_C=\frac{\hat{C}(m_0\vert\Omega\vert)^{\frac{\vert\hat{\Gamma}\vert}{2}}}{\sqrt{2\pi}}e^{i(m_j\varphi-Et)}\rho^{\vert\hat{\Gamma}\vert}e^{-\frac{m_0\vert\Omega\vert\rho^2}{2}}\left(
           \begin{array}{c}
            L^{\vert\hat{\Gamma}\vert}_{n}(m_0\vert\Omega\vert\rho^2) \\
            iD\left[\hat{R}(\rho)L^{\vert\hat{\Gamma}\vert}_{n}(m_0\vert\Omega\vert\rho^2)-2m_0\vert\Omega\vert\rho L^{\vert\hat{\Gamma}\vert+1}_{n-1}(m_0\vert\Omega\vert\rho^2)\right] \\
           \end{array}
         \right),
\end{equation}
where
\begin{equation}\label{D2}
D\equiv\frac{\tau}{(m_0+E-E_m)}, \ \ \hat{R}(\rho)\equiv\left((\vert\hat{\Gamma}\vert-s\hat{\Gamma})\frac{1}{\rho}-(\vert\Omega\vert+s\Omega)m_0\rho\right),
\end{equation}
with $\hat{C}$ being a new normalization constant.

Therefore, starting from the fact that $\Psi^{NC}_D=e^{-\frac{i\varphi\Sigma_3}{2}}\psi^{NC}_C$, we get the following NC Dirac spinor for the inertial curved case
\begin{equation}\label{spinor5} 
\Psi^{NC}_D(t,\rho,\varphi)=\hat{\Phi}(t,\rho,\varphi)\left(
           \begin{array}{c}
           L^{\vert\hat{\Gamma}\vert}_{n}(m_0\vert\Omega\vert\rho^2) \\
            iD\left[\hat{R}(\rho)L^{\vert\hat{\Gamma}\vert}_{n}(m_0\vert\Omega\vert\rho^2)-2m_0\vert\Omega\vert\rho L^{\vert\hat{\Gamma}\vert+1}_{n-1}(m_0\vert\Omega\vert\rho^2)\right] \\
           \end{array}
         \right),
\end{equation}
where
\begin{equation}\label{Phi2} 
\hat{\Phi}(t,\rho,\varphi)\equiv\frac{\hat{C}(m_0\vert\Omega\vert)^{\frac{\vert\hat{\Gamma}\vert}{2}}}{\sqrt{2\pi}}e^{i(m_l\varphi-Et)}\rho^{\vert\hat{\Gamma}\vert}e^{-\frac{m_0\vert\Omega\vert\rho^2}{2}}, \ \ \left(m_l=m_j\mp\frac{s}{2}\right).
\end{equation}

In particular, in the absence of the spinning CS ($\alpha=1$ and $\beta=0$) we obtain exactly the Dirac spinor in the Minkowski spacetime, given by (\ref{spinor3}).

\subsection{Nonrelativistic limit}\label{subsec2}

To analyze the nonrelativistic limit of our results, and in particular the relativistic spectrum, it is necessary to use the same standard prescription as in section 3.2, namely: $E\simeq m_0+\varepsilon$, where $m_0\gg\varepsilon$ and $m_0\gg E_m$. Therefore, using this prescription in (\ref{C7}) or on the spectrum (\ref{spectrum3}) for $l>0$ and $l<0$ with $s'=\pm 1$, we get the following nonrelativistic spectrum (nonrelativistic Landal levels) for the NCQHE with AMM for a spin-1/2 particle in the spinning CS spacetime (or in the presence of a declination)
\begin{equation}\label{spectrum4}
\varepsilon_{n,m_j,s}=E_{\alpha}+E_m+\vert\tau\lambda\vert\omega_c N_\alpha,
\end{equation}
where
\begin{equation}\label{E}
E_{\alpha}=s'm_0\omega_c\vert\lambda\vert\beta\left(\frac{\vert l \vert+s'l}{2\alpha l}\right)\gtrless 0, \ N_\alpha=\left(n+\frac{1}{2}+\frac{\vert l \vert+s'l+ss'\alpha}{2\alpha}\right)\geq 0,
\end{equation}
being $l=l(\alpha)=m_j-s\alpha/2$.

As we see in (\ref{spectrum4}), the nonrelativistic spectrum depends on the total magnetic quantum number $m_j=\pm\frac{1}{2},\pm\frac{3}{2},\ldots$, thus implying that such a spectrum is for a particle with spin up or down ($m_s=s/2$), and therefore, the equation of motion (or nonrelativistic wave equation) that describes the NCQHE must be something Pauli-like. In fact, this happens because there is no way to form an integer from the quantum number $l$ ($\alpha$ ``gets in the way'' of that goal). In particular, we note that the spectrum (\ref{spectrum4}) has some similarities and some differences with the relativistic case (for the particle). For example, unlike the relativistic case, the spectrum (\ref{spectrum4}) only admits (must, as they are nonrelativistic Landau levels) positive energy states ($\varepsilon_{n,m_l,s,s'}>0$), and depends linearly on the magnetic energy $E_m$, and also on the topological energy $E_{\alpha}$. Now, similar to the relativistic case, the spectrum (\ref{spectrum4}) also has its degeneracy broken (due to $\alpha$), increases as a function of quantum numbers $n$ and $m_j$ (for $s'l>0$) and of the parameters $\alpha$ and $\beta$, and still remains quantized even in the absence of the field, whose spectrum is given by: $\varepsilon_{n,l,s}=(\eta/m_0)\left[n+\frac{1}{2}+\frac{\vert l \vert+l+s\alpha}{2\alpha}\right]>0$. In particular, in the absence of the spinning CS ($\alpha=1$ and $\beta=0$) with $s=+1$ (spinless), we obtain exactly the usual spectrum of the NCQHE in Euclidean space, given by (\ref{spectrum2}). On the other hand, comparing the spectrum (\ref{spectrum4}) with the literature, we verified that even in the absence of the NC phase space ($\theta=\eta=0$), of the AMM ($E_m=0$), and considering only a static CS ($\beta=0$) with $s=+1$ (spinless), we do not obtain the spectrum of the QHE described by SE \cite{Lima}. In fact, this happens because even taking $s=+1$, there is no way to obtain the quantum number $m_l$, i.e., taking $s=+1$ we are just fixing the spin of the particle (spin up).

\section{The noncommutative Dirac equation in a rotating frame in the spinning cosmic string spacetime\label{sec6}}

To build the line element for a rotating frame in the spinning CS spacetime, in which we call from here the general background, we need to make a change in the angular coordinate as: $\varphi\to \varphi+\omega t$, where $\omega >0$ is the constant angular velocity of the rotating frame. Therefore, by making this change in the line element (\ref{CCgirante}), we get the following line element for our general background
\begin{equation}\label{generalbackground}
d\bar{s}_{CS}^2=(b^2-V^2)\left(dt+\frac{b\beta-V\alpha\rho}{(b^2-V^2)}d\varphi\right)^2-d\rho^2-\frac{\alpha^2\rho^2}{(b^2-V^2)}d\varphi^2,
\end{equation}
where the dimensionless parameter $b$ is defined as $b\equiv 1+\beta\omega>0$ and $V\equiv\alpha\omega\rho$ (SI: $V=\frac{\alpha\omega\rho}{c}=\frac{v}{c}$) is the ratio between the velocities of the rotating frame and the of light and must satisfy $V<b$ (causality requirement). In particular, taking $\omega=0$ in (\ref{generalbackground}) we have the line element of the spinning CS spacetime; taking $\omega=\beta=0$ and $\alpha=1$ we have we have the line element of the Minkowski spacetime; and taking only $\beta=0$, we have the line element of a rotating frame in the static CS spacetime \cite{OliveiraGRG,GRG,Oliveira2019,BakkePRD}. Furthermore, it is important to mention that the line element (\ref{generalbackground}) defines a new range for the radial coordinate $\rho$, given by: $0\leq\rho<\rho_0$, where $\rho_0\equiv\frac{b}{\alpha\omega}$ (SI: $\rho_0\equiv\frac{bc}{\alpha\omega}$). However, for all values where $\rho>\rho_0$ ($V>b$), it means that the fermion is outside the light cone, i.e., the speed of fermion is greater than that of light (something physically impossible). In this way, the interval $0\leq\rho<\rho_0$ imposes a spatial restriction where the Dirac spinor must be normalized, that is, that bound-state solutions (normalizable solutions) are reached. So, we must impose that such solutions disappear when $\rho\to\rho_0$ ($V=b$), with $\alpha\omega\ll 1$, which implies in $\rho_0\gg 1$, or ``$\rho_0 \to \infty$'' (a sufficiently large ``radius''), as well as disappearing when $\rho\to 0$ (now we have the two boundary conditions well defined) \cite{BakkePRD,OliveiraGRG,Castro,Matsuo1,Matsuo2,Chernodub1,Chernodub2}. Therefore, in this case where $ \alpha\omega \ll 1 $ we have $v=\alpha\omega\rho\to const. \neq 0 $, and consequently, $v\ll c$ or $V\ll b$ (no time dilation due to rotating frame).

So, comparing (\ref{generalbackground}) with (\ref{lineelement1}), we get the following coefficients
\begin{equation}\label{coefficients2}
X=\sqrt{b^2-V^2}, \ \ Y=\frac{b\beta-V\alpha\rho}{\sqrt{b^2-V^2}}, \ \ Z=\frac{\alpha\rho}{\sqrt{b^2-V^2}},
\end{equation}
and substituting (\ref{coefficients2}) in (\ref{dirac9}), we have the following NCDE in the general background
\begin{eqnarray}\label{CC1}
    \left\{\frac{1}{\sqrt{b^2-V^2}}\left[\gamma^0-\left(\frac{b\beta-V\alpha\rho}{\alpha\rho}\right)\gamma^2\right] E+i\tau\gamma^1\partial_\rho-\frac{\sqrt{b^2-V^2}}{\alpha\rho}\tau m_j\gamma^2-\frac{\lambda m_0\omega_c}{2}\rho\gamma^2\right\}\phi^{NC}
 \nonumber \\
    \label{eq:temporal-difference}
    +\left[i\chi\gamma^{1}+i\Lambda\gamma^{0}\gamma^1\gamma^2-\mu_m\Sigma^3 B-m_0\right]\phi^{NC}=0,
\end{eqnarray}
where
\begin{eqnarray}\label{CC2}
\chi=\frac{\sqrt{b^2-V^2}\left[(1-\tau) \left(\frac{V\alpha\omega\left(b \beta-V\alpha\rho\right)^2}{\left(b^2-V^2\right)^{3/2}}+\left(\frac{V\alpha\omega\left(b\beta-V\alpha\rho\right)}{\left(b^2-V^2\right)^{3/2}}-\frac{2V\alpha}{\sqrt{b^2-V^2}}\right)\left(b\beta-V\alpha\rho\right)\right)\right]}{4\alpha^2\rho^2}
 \nonumber \\
    \label{eq:temporal-difference}
    +\frac{\sqrt{b^2-V^2}\left[2\alpha\tau\rho\left(\frac{\alpha V^2}{\left(b^2-V^2\right)^{3/2}}+\frac{\alpha}{\sqrt{b^2-V^2}}\right)-\frac{2V^3\alpha^2\rho}{\left(b^2-V^2\right)^{3/2}}\right]}{4\alpha^2\rho^2},
\end{eqnarray}
and
\begin{equation}\label{CC3}
\Lambda=\frac{(2\tau-1)\sqrt{b^2-V^2} \left(\frac{V\alpha\omega\left(b\beta-V\alpha\rho\right)}{\left(b^2-V^2\right)^{3/2}}-\frac{2\alpha V}{\sqrt{b^2-V^2}}\right)+\frac{V\alpha\omega\left(b\beta-V\alpha\rho\right)}{b^2-V^2}}{4\alpha\rho}.
\end{equation}

However, we see that it is difficult to proceed without the simplification of Eq. \eqref{CC1}. Thus, to analytically solve Eq. \eqref{CC1}, we consider two approximations: the first is that the linear velocity of the rotation frame is much less than the speed of light ($V\ll b$), and the second is that the coupling between the angular momentum of the CS and the angular velocity of the rotating frame is very weak ($\beta\omega=\vec{\beta}\cdot\vec{\omega}\ll 1$) \cite{Muniz}. Therefore, using these two approximations in Eq. \eqref{CC1} with $b^{-1}\backsimeq(1-\beta\omega)$, we obtain 
\begin{eqnarray}\label{CC4}
    \left\{\frac{E}{b}\gamma^0+i\left(\tau\partial_\rho+\frac{\vec{\beta}\cdot\vec{\omega}(\tau-1)+\tau}{2\rho}\right)\gamma^1-\left[\frac{b\tau m_j+\beta E}{\alpha\rho}+\left(\frac{\lambda m_0\omega_c}{2}-\frac{\alpha\omega E}{b}\right)\rho\right]\gamma^2\right\}\phi^{NC}
 \nonumber \\
    \label{eq:temporal-difference}
    -\left[(2\tau-1)\alpha\gamma^{0}\vec{S}\cdot\vec{\omega}+\mu_m\Sigma^3 B+m_0\right]\phi^{NC}=0,
\end{eqnarray}
where the term $\vec{S}\cdot\vec{\omega}$ is called of spin-rotation coupling (of the fermion) \cite{Hehl,Matsuo1}, being $\vec{S}=\frac{1}{2}\vec{\Sigma}$, and $\vec{\omega}=\omega\vec{e}_z$, and we consider $(\alpha\omega)(\beta\omega)\approx 0$. Here, it is interesting to note that there are two kinds of spin-rotation coupling: that of the spinning CS with the rotating frame and that of the fermion with the rotating frame.

Now, using the form of the gamma matrices $\gamma^a=\tilde{\eta}^{ab}\gamma_{b}$ and the matrix $\Sigma^3=\Sigma_3$ as well as the spinor $\phi^{NC}$, implies that these two coupled differential equations are given by
\begin{equation}\label{CC5}
\frac{\left[\left(m_0+\frac{(2\tau-1)}{2}\alpha\omega\right)+\left(E_m-\frac{E}{b}\right)\right]}{\tau}f(\rho)=\left[\frac{d}{d\rho}+sm_0\bar{\Omega}\rho+\frac{s}{\rho}\left(\bar{m}_j+\frac{S}{2}\right)\right]g(\rho),
\end{equation}
\begin{equation}\label{CC6}
\frac{\left[\left(m_0+\frac{(2\tau-1)}{2}\alpha\omega\right)-\left(E_m-\frac{E}{b}\right)\right]}{\tau}g(\rho)=\left[\frac{d}{d\rho}-sm_0\bar{\Omega}\rho-\frac{s}{\rho}\left(\bar{m}_j-\frac{S}{2}\right)\right]f(\rho),
\end{equation}
where $\bar{m}_j \equiv \frac{1}{\alpha}(b m_j+\frac{\beta E}{\tau})$ is a new ``effective total magnetic quantum number'', $S\equiv s\left(1+\frac{\beta\omega(\tau-1)}{\tau}\right)$ is an ``effective spin parameter'', $\bar{\Omega}=\bar{\Omega}_{eff}\equiv(\frac{\lambda\omega_c}{2\tau}-\frac{\alpha\omega E}{m_0 b\tau})$ is a ``topological effective angular frequency'' (as it depends on $\alpha$), and $E_m=\mu_m B$ is the well-known magnetic energy.

Therefore, substituting (\ref{CC6}) into (\ref{CC5}), we obtain the following second-order differential DE (``quadratic DE'') for the NCQHE with AMM in a rotating frame in the spinning CS spacetime
\begin{equation}\label{CC7}
\left[\frac{d^2}{d\rho^2}+\frac{sS}{\rho}\frac{d}{d\rho}-\frac{\bar{\bar{\Gamma}}^2}{\rho^2}-(m_0\bar{\Omega}\rho)^2+\bar{\bar{\mathcal{E}}}\right]f(\rho)=0,
\end{equation}
where define
\begin{equation}\label{CC8}
\bar{\bar{\Gamma}}\equiv\sqrt{\left(\bar{m}_j-\frac{s}{2}\right)^2-\frac{1}{4}(1-sS)^2},
\end{equation}
and
\begin{equation}\label{CC9}
\bar{\bar{\mathcal{E}}}\equiv\frac{\left(\frac{E}{b}-E_m\right)^2-\left(m_0+\frac{(2\tau-1)}{2}\alpha\omega\right)^2}{\tau^2}-2m_0\bar{\Omega}\left(\bar{m}_j+\frac{s}{2}\right).
\end{equation}

In particular, in the absence of the  rotating frame ($\omega=0$) we obtain exactly the NCDE in the spinning CS spacetime, given by (\ref{C4}).

\subsection{Bound-state solutions: two-component Dirac spinor and relativistic Landau levels}\label{subsec1}

To analytically solve Eq. (\ref{CC7}), let's follow the same procedure done in section 3.1 (here it is not enough to just replace $m_j$ with $\bar{m}_j$). Therefore, considering for simplicity that $\tau>0$ and $\lambda>0$, and introducing a new (dimensionless) variable given by: $\bar{r}=m_0\vert\bar{\Omega}\vert\rho^2>0$ ($\vert\bar{\Omega}\vert=\bar{\Omega}$), Eq. \eqref{CC7} becomes
\begin{equation}\label{CC10}
\left[\bar{r}\frac{d^{2}}{d\bar{r}^{2}}+\frac{(sS+1)}{2}\frac{d}{d\bar{r}}-\frac{\bar{\bar{\Gamma}}^{2}}{4\bar{r}}-\frac{\bar{r}}{4}+\frac{\bar{\bar{\mathcal{E}}}}{4m_0\vert\bar{\Omega}\vert}\right]f(\bar{r})=0.
\end{equation}

Now, analyzing the asymptotic behavior of Eq. (\ref{CC10}) for $\bar{r}\to 0$ and $\bar{r}\to\infty$, we get a (regular) solution to this equation given by the following ansatz
\begin{equation}\label{CC11}
f(\bar{r})=\bar{\bar{C}}\bar{r}^{\chi}e^{-\frac{\bar{r}}{2}}F(\bar{r}),
\end{equation}
where $\chi\equiv \frac{2\vert\bar{m}_j-s/2\vert+(1-sS)}{4}$, $\bar{C}>0$ is a new normalization constant, and $F(\bar{r})$ is an unknown function to be determined.

So, substituting \eqref{CC11} in \eqref{CC10}, we have a differential equation for $F(\bar{r})$ as
\begin{equation}\label{CC12}
\left[\bar{r}\frac{d^{2}}{d\bar{r}^{2}}+(\Delta-\bar{r})\frac{d}{d\bar{r}}-\bar{\Delta}\right]F(\bar{r})=0,
\end{equation}
with
\begin{equation}\label{CC13}
\Delta\equiv\Big\vert\bar{m}_j-\frac{s}{2}\Big\vert+1, \ \ \bar{\Delta}\equiv\frac{\Delta}{2}-\frac{\bar{\bar{\mathcal{E}}}}{4m_0\vert\bar{\Omega}\vert}.
\end{equation}

According to the literature \cite{Villalba}, Eq. \eqref{CC12} is an associated Laguerre equation, whose solution are the associated Laguerre polynomials $F(\bar{r})=L^{\vert\bar{m}_j-\frac{s}{2}\vert}_n(\bar{r})$. Consequently, the quantity $\bar{\Delta}$ must to be equal to a non-positive integer, i.e., $\bar{\Delta}=-n$ (quantization condition), where $n=n_\rho=0,1,2,\ldots$ is the radial quantum number. Therefore, from this condition, we have the following quadratic polynomial equation for the relativistic total energy $E$
\begin{equation}\label{CC14}
E^2-[2bE_m-2b\tau\alpha\omega\bar{n}]E-\frac{2m_0 b^2}{\alpha}\left(\frac{\tau\lambda\omega_c}{2}-\frac{\tau\alpha\omega E}{m_0 b}\right)\left[\left(\bar{l}+\frac{\beta E}{\tau}\right)+\Big\vert\bar{l}+\frac{\beta E}{\tau}\Big\vert\right]+[b^2E^2_m-b^2 M^2_0-b^2\tau\lambda m_0\omega_c\bar{n}]=0,
\end{equation}
where
\begin{equation}\label{CC15}
M_0\equiv\left(m_0+\frac{(2\tau-1)}{2}\alpha\omega\right),
\end{equation}
being $\bar{l}\equiv bm_j-\frac{s\alpha}{2}$ is a new ``topological quantum number'', and we define for simplicity that $\bar{n}\equiv 2n+1+s$.

Therefore, analyzing the polynomial equation (\ref{CC14}) for $\bar{l}>0$ and $\bar{l}<0$, we obtain the following relativistic spectrum for the NCQHE with AMM in a rotating frame in the spinning CS spacetime
\begin{equation}\label{spectrum5}
E^{\kappa}_{n,l,s}=\frac{\left[\bar{E}_\alpha+\bar{E}_m-E_\omega\right]}{\left[1+4\beta\omega\left(\frac{\vert\bar{l}\vert+\bar{l}}{2\bar{l}}\right)\right]}+\kappa\sqrt{\frac{\left[\bar{E}_\alpha+\bar{E}_m-E_\omega\right]^2}{\left[1+4\beta\omega\left(\frac{\vert\bar{l}\vert+\bar{l}}{2\bar{l}}\right)\right]^2}+\frac{[-\bar{E}^2_m +\bar{M}_0^2+2\bar{b}\tau\lambda m_0\omega_c \bar{N}_{\alpha}]}{\left[1+4\beta\omega\left(\frac{\vert\bar{l}\vert+\bar{l}}{2\bar{l}}\right)\right]}},
\end{equation}
where
\begin{equation}\label{N1}
\bar{E}_\alpha\equiv \bar{b}m_0\omega_c\tau\lambda\beta\left(\frac{\vert\bar{l}\vert+\bar{l}}{2\alpha\bar{l}}\right)>0, \ \ \bar{E}_m \equiv b E_m>0, \ \ E_\omega \equiv 2\alpha\omega\bar{N}_\alpha^\tau\geq 0, \  \ \bar{M}_0\equiv b M_0>0,
\end{equation}
and
\begin{equation}\label{N2}
\bar{N}_\alpha\equiv\left(\frac{\bar{n}}{2}+\frac{\vert\bar{l}\vert+\bar{l}}{2\alpha}\right)\geq 0, \ \ \bar{N}_\alpha^\tau\equiv\left(\tau\frac{\bar{n}}{2}+\frac{\vert\bar{l}\vert+\bar{l}}{2\alpha}\right)\geq 0,
\end{equation}
being $\kappa=\pm 1$ the known well energy parameter (particle/antiparticle), $E_\omega$ is a kind of quantized rotational energy (analogous to the rotational spectrum of diatomic molecules modeled by a rigid rotor) or a ``topological rotational energy'' (because it also depends on $\alpha$), and the parameter $\bar{b}$ is given by $\bar{b}=1+2\beta\omega$ (arises from $b^2$ with $\beta\omega\ll 1$). In particular, in the absence of the rotating frame ($\omega=0$), we obtain exactly (with $s'=+1$) the spectrum in the spinning CS spacetime, given by (\ref{spectrum3}). However, unlike to the spectrum (\ref{spectrum3}), the spectrum (\ref{spectrum4}) still depends on $\alpha$ and $\beta$ even for $\bar{l}<0$. Indeed, this happens due to the presence of the angular velocity $\omega$, where $\omega$ is ``tied'' to the parameter $\beta$ in $b$ and $\bar{b}$, and ``tied'' to the parameter $\alpha$ in $E_\omega$. Besides, unlike to the spectrum (\ref{spectrum3}), the spectrum (\ref{spectrum4}) still remains quantized even in the absence of the magnetic field ($B=0$) and of the NC of the momentum ($\eta=0$), where we have $E_\omega$. On the other hand, analogous to the spectrum (\ref{spectrum3}), the degeneracy of the Landau levels still remains broken here, and the energies of the particle and antiparticle are larger when both have spin up ($s=+1$).

\begin{figure}[ht]
\centering
\includegraphics[scale=1.0]{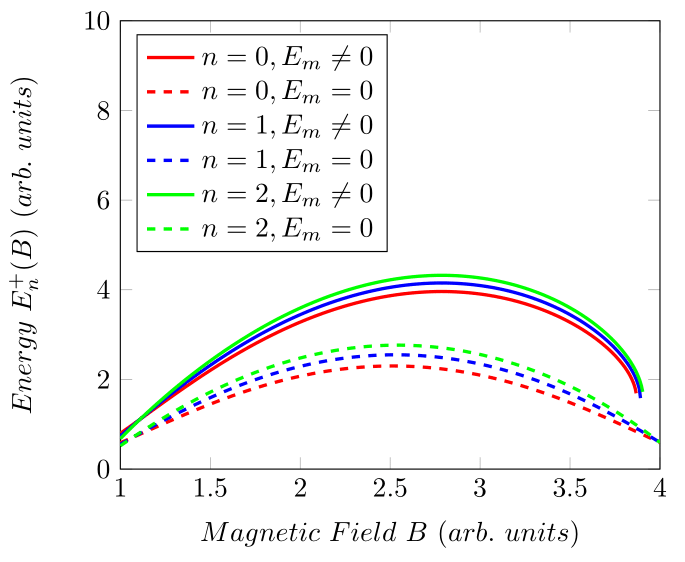}
\caption{Graph of $E^+_n(B)$ versus $B$ for three different values of $n$ with $E_m\neq 0$ ($a=1$) and $E_m=0$ ($a=0$).}
\label{fig9}
\end{figure}

Now, let's analyze the behavior of the spectrum (\ref{spectrum5}) as a function of the magnetic field and of the angular velocity $\omega$ for different values of $n$ (with $m_j$ fixes). Therefore, using the restriction 1 of Table \ref{tab3} we have Fig. \ref{fig9}, where it shows the behavior of the energies of the particle as a function of the magnetic field for the ground state ($n=0$) and the first two excited states ($n=1,2$), with and without the presence of magnetic energy, in which we take $m_0=e=a=\theta=\eta=1$, $\beta=\omega=0.1$, $\alpha=1/2$ (must satisfy $\beta\omega\ll 1$ and $\alpha\omega\ll 1$), $m_j=1/2$, and $s=+1$ (spin up), being the field given by $1<B<4$. According to this figure, we see that the energies increase with the increase of $n$ (as it should be) and the function of the AMM is to increase the energies of the particle (analogous to the two previous cases). Also, energies can increase or decrease as a function of $B$ (analogous to the inertial flat case). For example, for the case $E_m\neq 0$ the energies increase between $B\approx 1$ and $B\approx 2.8$, and decrease between $B\approx 2.8$ and $B\approx 3.8$ (with $\Delta E=E_{final}-E_{initial}>0$), while for the case $E_m=0$ the energies increase between $B\approx 1$ and $B=2.5$, and decrease between $B=2.5$ and $B\approx 4$ ($\Delta E=0$). However, unlike the inertial flat case, here, the maximum energies for the particle (in $B\approx 2.8$) and for the antiparticle (in $B=2.5$) are higher.

\begin{figure}[ht]
\centering
\includegraphics[scale=1.0]{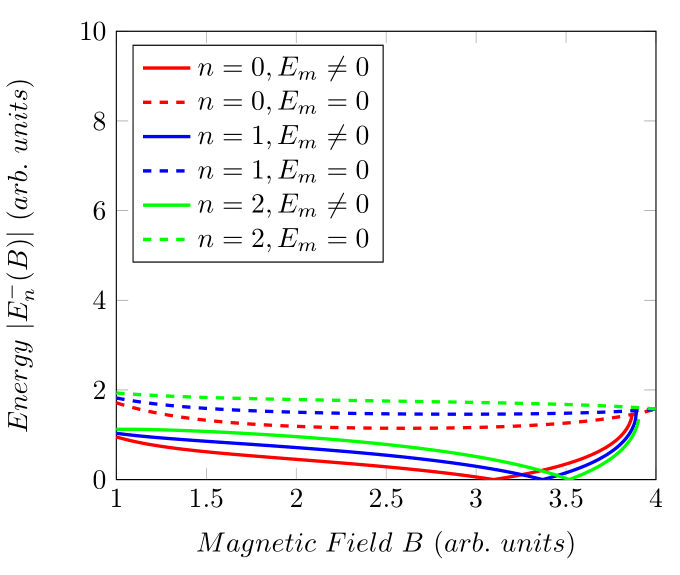}
\caption{Graph of $\vert E^{-}_n (B)\vert$ versus $B$ for three different values of $n$ with $E_m\neq 0$ ($a=1$) and $E_m=0$ ($a=0$).}
\label{fig10}
\end{figure}

In Fig. \ref{fig10}, we see the behavior of the energies of the antiparticle as a function of the magnetic field for the ground state and the first two excited states (with $m_0=e=a=\theta=\eta=1$, $\beta=\omega=0.1$, $\alpha=1/2$, $m_j=1/2$, $s=+1$, and $1<B<4$). According to this figure, we see that the energies for the case $E_m=0$ are not equal to those of the particle also with $E_m=0$ (asymmetric spectra). In particular, this behavior is different from the inertial flat case but analogous to the inertial curved case. Indeed, the presence of a spinning CS (not static) or a noninertial frame can cause the symmetry of the spectra to break. Now, analogous to the inertial flat case (but unlike the inertial curved case), the energies of the antiparticle are larger in the absence of magnetic energy and can be ``null'' for certain values of the magnetic field. Then, solving $\vert E^{-}_{n}(B)\vert=0$ for each specific state, we have $B\approx 3.1$ ($n=0$), $B\approx 3.4$ ($n=1$), and $B\approx 3.5$ ($n=2$), respectively. Therefore, we see that these values of $B$ are very close to the inertial flat case.

\begin{figure}[ht]
\centering
\includegraphics[scale=1.0]{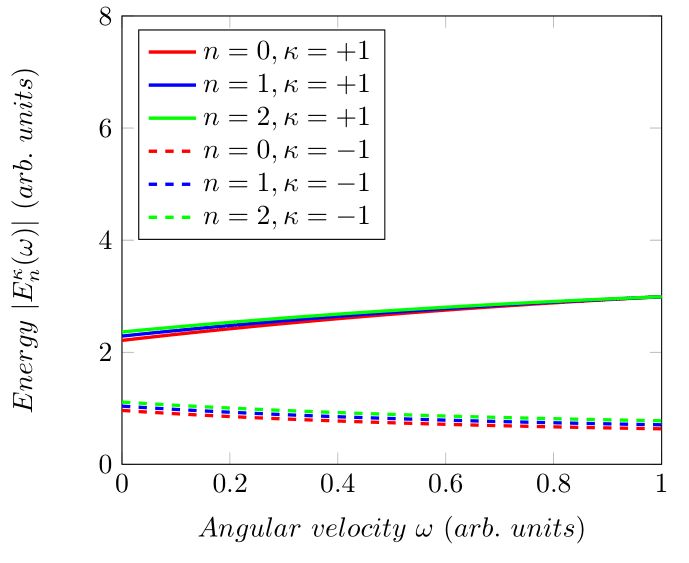}
\caption{Graph of $\vert E^{\kappa}_n(\omega)\vert$ versus $\omega$ for three different values of $n$.}
\label{fig11}
\end{figure}

Now, let's analyze the behavior of the spectrum (\ref{spectrum5}) as a function of the angular velocity $\omega$ for different values of $n$ (with $m_j$ fixes). Therefore, using the restriction 1 of Table \ref{tab3} and also the restriction 2 of Table \ref{tab5}, we have Fig. \ref{fig11}, where it shows the behavior of the energies of the particle and antiparticle as a function of $\omega$ for the ground state ($n=0$) and the first two excited states ($n=1,2$), in which we take $m_0=e=a=B=1$, $\theta=3$, $\eta=0.5$, $\beta=\alpha=0.1$ (must satisfy $\beta\omega\ll 1$ and $\alpha\omega\ll 1$), $m_j=1/2$, and $s=+1$. According to this figure, we see that the energies increase with the increase of $n$ (as it should be); however, the energy difference between one level and another is very small. In particular, in the case of the particle, this energy difference is practically zero when $\omega\to 1$. Also, the function of $\omega$ is to increase the energies of the particle and decrease the energies of the antiparticle, where the energies of the particle are always greater than those of the antiparticle.

Before ending this section, let's now focus on the form of the NC Dirac spinor (noninertial curved spinor) for the relativistic bound states of the NCQHE, given by $\Psi^{NC}_D=U\psi^{NC}_C$. So, using the fact that the dimensionless variable $\bar{r}$ is written as $\bar{r}=m_0\vert\bar{\Omega}\vert\rho^2$, implies that we can rewrite the function \eqref{CC11} as follows
\begin{equation}\label{CC16}
f(\rho)=\bar{C}(m_0\vert\bar{\Omega}\vert)^{\chi}\rho^{2\chi}e^{-\frac{m_0\vert\bar{\Omega}\vert\rho^2}{2}}L^{\vert\bar{m}_j-\frac{s}{2}\vert}_n(m_0\vert\bar{\Omega}\vert\rho^2).
\end{equation}

On the other hand, to obtain the inferior radial function of the spinor, given by $g(\rho)$, just substitute (\ref{CC16}) in Eq. (\ref{CC6}), where we get
\begin{equation}\label{CC17}
g(\rho)=\bar{C}\bar{D}(m_0\vert\bar{\Omega}\vert)^{\chi}\rho^{2\chi}e^{-\frac{m_0\vert\bar{\Omega}\vert\rho^2}{2}}\left[\bar{R}(\rho)L^{\vert\bar{m}_j-\frac{s}{2}\vert}_{n}(m_0\vert\bar{\Omega}\vert\rho^2)-2m_0\vert\bar{\Omega}\vert\rho L^{\vert\bar{m}_j-\frac{s}{2}\vert+1}_{n-1}(m_0\vert\bar{\Omega}\vert\rho^2)\right],
\end{equation}
with
\begin{equation}\label{D3}
\bar{D}\equiv\frac{\tau}{\left[\left(m_0+\frac{(2\tau-1)}{2}\alpha\omega\right)+\frac{E}{b}-E_m\right]}, \ \ \bar{R}(\rho)\equiv\left\{\left[2\chi-s\left(\bar{m}_j-\frac{S}{2}\right)\right]\frac{1}{\rho}-(1+s)m_0\vert\bar{\Omega}\vert\rho\right\}.
\end{equation}

Then, from the radial functions (\ref{CC16}) and (\ref{CC17}), it implies that the curvilinear spinor  for the noninertial curved case takes the following form
\begin{equation}\label{spinor6}
\psi^{NC}_C=\frac{\bar{C}(m_0\vert\bar{\Omega}\vert)^{\chi}}{\sqrt{2\pi}}e^{i(m_j\varphi-Et)}\rho^{2\chi}e^{-\frac{m_0\vert\bar{\Omega}\vert\rho^2}{2}}\left(
           \begin{array}{c}
            L^{\vert\bar{m}_j-\frac{s}{2}\vert}_{n}(m_0\vert\bar{\Omega}\vert\rho^2) \\
            i\bar{D}\left[\bar{R}(\rho)L^{\vert\bar{m}_j-\frac{s}{2}\vert}_{n}(m_0\vert\bar{\Omega}\vert\rho^2)-2m_0\vert\bar{\Omega}\vert\rho L^{\vert\bar{m}_j-\frac{s}{2}\vert+1}_{n-1}(m_0\vert\bar{\Omega}\vert\rho^2)\right] \\
           \end{array}
         \right).
\end{equation}

Therefore, as $\Psi^{NC}_D=e^{-\frac{i\varphi\Sigma_3}{2}}\psi^{NC}_C$, we get the following NC Dirac spinor for the noninertial curved case
\begin{equation}\label{spinor7} 
\Psi^{NC}_D(t,\rho,\varphi)=\bar{\Phi}(t,\rho,\varphi)\left(
           \begin{array}{c}
            L^{\vert\bar{m}_j-\frac{s}{2}\vert}_{n}(m_0\vert\bar{\Omega}\vert\rho^2) \\
            i\bar{D}\left[\bar{R}(\rho)L^{\vert\bar{m}_j-\frac{s}{2}\vert}_{n}(m_0\vert\bar{\Omega}\vert\rho^2)-2m_0\vert\bar{\Omega}\vert\rho L^{\vert\bar{m}_j-\frac{s}{2}\vert+1}_{n-1}(m_0\vert\bar{\Omega}\vert\rho^2)\right] \\
           \end{array}
         \right),
\end{equation}
where
\begin{equation}\label{Phi3} 
\bar{\Phi}(t,\rho,\varphi)\equiv\frac{\bar{C}(m_0\vert\bar{\Omega}\vert)^{\chi}}{\sqrt{2\pi}}e^{i(m_l\varphi-Et)}\rho^{2\chi}e^{-\frac{m_0\vert\bar{\Omega}\vert\rho^2}{2}}, \ \ \left(m_l=m_j\mp\frac{s}{2}\right).
\end{equation}

In particular, in the absence of the rotating frame ($\omega=0$) we obtain exactly the Dirac spinor in the spinning CS spacetime, given by (\ref{spinor6}).

\subsection{Nonrelativistic limit}\label{subsec2}

To analyze the nonrelativistic limit of our results, and in particular the relativistic spectrum, we can not just use the standard prescription from section 3.2, where the objective was to cancel quadratic terms of $m_0$ and $E_m$. Here, we must make a correction in this standard prescription due to the rotating frame, i.e., we consider that: $E\simeq b(m_0+\varepsilon)$, where $m_0\gg\varepsilon$, $m_0\gg E_m$ and $m_0\gg \alpha\omega$. Therefore, using this prescription in (\ref{CC14}) or on the spectrum (\ref{spectrum5}) for $l>0$ and $l<0$, we get the following nonrelativistic spectrum (nonrelativistic Landal levels) for the NCQHE with AMM for a spin-1/2 particle in a rotating frame in the spinning CS spacetime (or in the presence of a declination)
\begin{equation}\label{spectrum6}
\varepsilon_{n,l,s}=\frac{\left[\bar{E}_\alpha+\bar{E}_m-E_\omega+b\tau\lambda\omega_c \bar{N}_{\alpha}+\frac{(2\tau-1)}{2}\alpha\omega\right]}{\left[1+4\beta\omega\left(\frac{\vert\bar{l}\vert+\bar{l}}{2\bar{l}}\right)\right]},
\end{equation}
where
\begin{equation}\label{N3}
\bar{E}_\alpha\equiv \bar{b}m_0\omega_c\tau\lambda\beta\left(\frac{\vert\bar{l}\vert+\bar{l}}{2\alpha\bar{l}}\right)>0, \ \ \bar{E}_m \equiv b E_m>0, \ \ E_\omega \equiv 2\alpha\omega\bar{N}_\alpha^\tau\geq 0,
\end{equation}
and
\begin{equation}\label{N4}
\bar{N}_\alpha\equiv\left(\frac{\bar{n}}{2}+\frac{\vert\bar{l}\vert+\bar{l}}{2\alpha}\right)\geq 0, \ \ \bar{N}_\alpha^\tau\equiv\left(\tau\frac{\bar{n}}{2}+\frac{\vert\bar{l}\vert+\bar{l}}{2\alpha}\right)\geq 0.
\end{equation}

In particular, in the absence of the rotating frame ($\omega=0$), we obtain exactly (with $s'=+1$) the nonrelativistic spectrum in the spinning CS spacetime, given by \eqref{spectrum4}. Now, ignoring the last term in \eqref{spectrum6} and in the absence of the NC phase space ($\theta=\eta=0$), AMM ($E_m=0$), spinning CS ($\alpha=1$ and $\beta$), with $m_l\leq 0$ and $s=+1$, we get the spectrum of the nonrelativistic QHE in a rotating frame for electrons \cite{Filgueiras}. In addition, we note that the spectrum (\ref{spectrum6}) has some similarities and some differences with the relativistic case (for the particle). For example, unlike the relativistic case, the spectrum (\ref{spectrum6}) only admits positive energy states ($\varepsilon_{n,l,s}>0$), and depends linearly on the magnetic energy $E_m$, topological energy $\bar{E}_{\alpha}$, and also on the rotational energy $E_\omega$. On the other hand, similar to the relativistic case, the spectrum (\ref{spectrum6}) also has its degeneracy broken (due to $\alpha$), increases as a function of quantum numbers $n$ e $m_j$ (for $l>0$), and still remains quantized even in the absence of the field or of the NC phase space (due to $E_\omega$).

\section{Conclusion}\label{sec6}

In this paper, we investigate the bound-state solutions of the NCQHE with AMM in three different relativistic scenarios, namely: the Minkowski spacetime (inertial flat case), the spinning CS spacetime (inertial curved case), and the spinning CS spacetime with noninertial effects (noninertial curved case). In particular, in the first two scenarios, we have an inertial frame, while in the third, we have a noninertial frame or rotating frame. With respect to bound-state solutions, we focus primarily on eigenfunctions (two-component Dirac spinor and Schr\"{o}dinger wave function) and on energy eigenvalues (energy spectrum or Landau levels), where we use the DE in polar coordinates to reach such solutions. However, unlike the literature, here we consider a CS with an angular momentum non-null and also the NC of the positions, i.e., we seek a more general description for the QHE. Furthermore, to analytically solve our equations also for the third case, we consider two approximations: the first is that the linear velocity of the rotating frame is much less than the speed of light, and the second is that the coupling between the angular momentum of the CS and the angular velocity of the rotating frame is very weak (weak spin-rotation coupling).

Our main results can be summarized as follows:

\begin{itemize}
\item $1\textsuperscript{\underline{o}}$ Scenario. Analyzing the asymptotic behavior of our differential equation, we determined the Dirac spinor and the relativistic energy spectrum, where we verified that such spinor is written in terms of the associated Laguerre polynomials and such spectrum linearly depends on the magnetic energy $E_m$ generated by the interaction of the AMM with the external magnetic field $B$, as well as explicitly depends on the quantum numbers $n$ and $m_j$, cyclotron frequency $\omega_c$, spin parameter $s$, and on the NC parameters $\theta$ and $\eta$. So, analyzing this spectrum in detail, we have the following conclusions: can be finitely or infinitely degenerate depending on the values of $s'$, $s$, and $m_j$, where the parameter $s’$ arises from restrictions on $B$, $\theta$ and $\eta$ (only specific values are allowed for $B$, $\theta$, and $\eta$); only increases as a function of $n$ and $m_j$ for $s'=+1$ with $m_j>0$, or $s'=-1$ with $m_j<0$; the energies of the particle are higher than those of the antiparticle (asymmetric spectra); the function of the AMM is to increase (decrease) the energies of the particle (antiparticle); the energies of the particle can increase or decrease as a function of $B$, while the energies of the antiparticle can increase, decrease, or be ``null''; the function of $\theta$ and $\eta$ is to increase the energies; and even in the absence of the magnetic field ($B=0$), the spectrum still remains quantized (due to $\eta$). In the nonrelativistic limit, we obtain the SE with two types of Hamiltonians: the quantum harmonic oscillator-like Hamiltonian and the Zeeman hamiltonian, or two types of MDMs: the orbital and the anomalous. Besides, unlike the relativistic case, the nonrelativistic spectrum only admits positive energies, and linearly depends on $n$, $m_l$, $\omega_c$, $\theta$ and $\eta$. Now, similar to the relativistic case, the nonrelativistic spectrum linearly depends on $E_m$, has a finite or infinite degeneracy, increases as a function of $n$ e $m_l$, and still remains quantized even in the absence of the magnetic field. On the other hand, comparing our results with other works, we verified that our energy spectra generalize several particular cases of the literature.

\item $2\textsuperscript{\underline{o}}$ Scenario. Comparing our differential equation with that of the $1\textsuperscript{\underline{o}}$ scenario, we note that both are very similar, consequently, the results of the $2\textsuperscript{\underline{o}}$ scenario can be obtained simply replace $\Gamma$ by $\hat{\Gamma}=\hat{m}_j-\frac{s}{2}$ and $m_j$ by $\hat{m}_j=\frac{1}{\alpha}(m_j+\frac{\beta E}{\tau})$, where $\alpha$ and $\beta$ are the topological and rotational parameters of the spinning CS. In that way, we determined the Dirac spinor and the relativistic spectrum for the inertial curved case, where such spectrum linearly depends on a ``topological energy'' $E_\alpha$ (only exists because of the magnetic field) and explicitly depends on $\alpha$ and $\beta$, and on the ``topological quantum number'' $l=l(\alpha)=m_j-\frac{s\alpha}{2}$. So, analyzing this spectrum in detail, we have the following conclusions: only depends on both $\alpha$, $\beta$ and $m_j$ for $l>0$ with $s'=+1$, or $l<0$ with $s'=-1$; the degeneracy of the spectrum is broken (due to $\alpha$), the function of the AMM is to increase both the energies of the particle and antiparticle; the energies of the particle increase as a function of $B$, while the energies of the antiparticle decrease; the function of $\alpha$ and $\beta$ is to increase the energies of the particle and decrease those of the antiparticle; and even in the absence of the magnetic field the spectrum still remains quantized. In the nonrelativistic limit, we also obtain the spectrum for a particle with spin-1/2 (Pauli-like particle) since such a spectrum still depends on the quantum number $m_j$. Besides, unlike the relativistic case, the nonrelativistic spectrum only admits positive energies and depends linearly on $E_m$ and $ E_{\alpha}$. Now, similar to the relativistic case, the nonrelativistic spectrum also has its degeneracy broken, increases as a function of $n$ and $m_j$, and still remains quantized even in the absence of the field. In particular, taking $\alpha\to 1$ and $\beta\to 0$ (absence of the spinning CS), we recover the results of the $1\textsuperscript{\underline{o}}$ scenario.

\item $3\textsuperscript{\underline{o}}$ Scenario. Analyzing the asymptotic behavior of our differential equation, we determined the Dirac spinor and the relativistic energy spectrum, where we verified that such spinor is also written in terms of the associated Laguerre polynomials and such spectrum linearly depends on the (quantized) rotational energy $E_\omega$ (analogous to the rotational spectrum of diatomic molecules), and explicitly depends on a new ``topological quantum number'', given by $\bar{l}=\bar{l}(\alpha)=bm_j-\frac{s\alpha}{2}$, where $b=1+\beta\omega$. So, analyzing this spectrum in detail, we have the following conclusions: still depends on $\alpha$ and $\beta$ even for $\bar{l}<0$; the function of the AMM and of the angular velocity $\omega$ is to increase (decrease) the energies of the particle (antiparticle); the energies of the particle can increase or decrease as a function of $B$, while the energies of the antiparticle can increase, decrease, or be ``null''; and still remains quantized even in the absence of the magnetic field ($B=0$) and of the NC of the momentum ($\eta=0$). In particular, taking $\omega\to 0$ (absence of the rotating frame), we recover the results of the $2\textsuperscript{\underline{o}}$ scenario. Now, in the absence of the NC phase space ($\theta=\eta=0$), AMM ($E_m=0$), spinning CS ($\alpha=1$ and $\beta$), with $m_l\leq 0$ and $s=+1$, we get the spectrum of the nonrelativistic QHE in a rotating frame. 
\end{itemize}

\section*{Acknowledgments}
The authors would like to thank the Funda\c c\~ao Cearense de Apoio ao Desenvolvimento
Cient\'ifico e Tecnol\'ogico (FUNCAP), the Coordena\c c\~ao de Aperfei\c c\ oamento de Pessoal de
N\'ivel Superior (CAPES), and the Conselho Nacional de Desenvolvimento Cient\'ifico e Tecnol\'ogico (CNPq) for financial support.

\end{document}